# The $p$-Innovation Ecosystems problem

**Daniel E. Restrepo [1,2], Juan C. Duque [2,✉], and Richard Church [3]**

[1] Department of Mathematics. The University of Texas at Austin. TX, USA.

[2] Research in Spatial Economics (RiSE) Group. Universidad EAFIT. Colombia.

[3] Department of Geography,University of California, Santa Barbara, USA.

✉Correspondence: jduquec1@eafit.edu.co. Carrera 49, 7 Sur-50. Department of Mathematical Sciences. Universidad EAFIT. 050022, Medellín, Antioquia, Colombia.

August 2, 2020

**Abstract**

In this paper, we propose a spatially constrained clustering problem belonging to the family of $p$-regions problems. Our formulation is motivated by the recent developments of economic complexity on the evolution of the economic output through key interactions among industries within economic regions. The objective of this model consists in aggregating a set of geographic areas into a prescribed number of regions (so-called innovation ecosystems) such that the resulting regions preserve the most relevant interactions among industries. We formulate the $p$-Innovation Ecosystems model as a mixed-integer programming (MIP) problem and propose a heuristic solution approach. We explore a case involving the municipalities of Colombia to illustrate how such a model can be applied and used for policy and regional development.

**funding:** This document was completed with support from the World Bank, Contract Ref.: 1029-000001, in addition to support from the PEAK Urban programme, supported by UKRIs Global Challenge Research Fund, Grant Ref.: ES/P011055/1

**Acknowledgments:** The authors are grateful for the support with the Apolo supercomputer resources and the team at EAFIT, specially to Sebastián Patiño. We are also grateful to Vanessa Echeverri for her constant support during the last stage of this project. The usual disclaimer applies.

**Conflicts of Interest:** The authors declare no conflict of interest.

# The $p$-Innovation Ecosystems problem.

## 1 Introduction

Since the seminal papers of Robert Solow, the mainstream analyses of economic growth have had aggregate growth models (Solow, 1957; Solow, 1956) as its principal workhorse. Abstract quantities such as technological progress or positive externalities, like spillovers or scale effects, usually play a central role in classical growth theory since they are, at last, the primary mechanisms that encourage firms productivity and, therefore, generate economic growth. Nevertheless, given that the aggregative models address the interaction between different goods as a black box, they cannot endogenize the positive externalities that come from the strategic production of complementary goods; like the scale effects or the spillover effects mentioned above (Hidalgo et al., 2007).

Regardless of the approach (aggregative or not), the general consensus is that the diversification of the productive structure (i.e., selective innovation) stands out as a fundamental goal to foster economic growth. Thus, many theoretical efforts have been directed to understand and identify suitable environments for innovation such as cities or industrial clusters (see e.g. Krugman, 1991). Classic examples of such ideas within the context of urban economics are the so-called Jacobs' externalities (Jacobs, 2016) and the agglomeration and location externalities of the New Economic Geography (NEG). The approaches of Jacobs and the NEG intended to provide a more detailed study of the positive effects of industrial interaction, advocating for the diversification of the production and the division of labor as concomitant processes that boost innovation opportunities.

Understanding the relatedness structure among economic activities thus becomes a key part of the analysis of diversification and agglomeration of firms. Several studies following the pioneering ideas introduced by Hausmann and Hidalgo have found in network theory a suitable framework to understand the impacts of the productive structure of countries and regions on their diversification patterns (Hidalgo et al., 2007; Neffke et al., 2011; Hausmann and Hidalgo, 2011; Hausmann et al., 2014). Hidalgo et al. (2007) introduced the Product Space (PS), an undirected network capturing the relatedness structure of the tradable goods of the world market. This network revealed a core-periphery distribution, where the more complex goods ,i.e., the ones requiring more know-how, are highly connected whereas goods requiring fewer know-how and skills are more likely to be leaves of the graph. Complex goods are present in countries with a diverse export basket and according to Hidalgo et al. (2007) and Hausmann and Hidalgo (2011) arise naturally by the interaction of existing industries. Therefore, enhancing the production of complex goods and fostering industrial interactions that lead to their emergence are natural policy goals to pursue. In fact, Hausmann et al. (2014) shows that the amount of complexity reached by a country is a strong predictor of its economic growth.

However, geographic, institutional and historical factors such as distance among firms, topography, administrative borders, differential taxes among states, productivity, labor availability and so forth are important constraints that may preclude the formation of optimal industrial clusters, i.e., the idiosyncratic factors of the territory can encourage the agglomeration of industries producing peripheral, or less valuable, industries in terms of the PS. This clustering process, although suboptimal, ends up reinforcing itself by the impact of the agglomeration economies that, despite all the possible drawbacks like congestion or elevated factor costs, leads the firms to locate close to each other (Diodato et al., 2018 and Rosenthal and Strange, 2004). On top of that, empirical studies have shown that the predictions of NEG (Martin and Sunley, 1996) are not robust and that the effects of Jacobs externalities are (at most) inconclusive (De Groot et al., 2016). It can be argued that the lack of conclusive empirical validation of the aforementioned theories is partially explained by the trade-off between diversity and similarity (see Neffke et al., 2011): labor, knowledge, and know-how diffuse easily among similar firms, but only the interaction between firms distinct enough can induce the generation of new ideas. More precisely, this idea (taken from cognitive theory) explain how a non-directed diversification can either lead to big dissimilarities among firms (cognitive distance) hampering interaction or lead to an excessive overlapping of skills amounting to cognitive lock-in (Neffke et al., 2011). In the case of regions, this trade-off can be put in terms of capacity of a region to incorporate technologies or knowledge from other region (absorptive capacity) (Jung and López-Bazo, 2017). This concept has turned out to be useful to explain disparities in productivity levels among economic regions and, analogously, to explain the cognitive lock-in among firms. It has been noticed that some kinds of poverty traps are generated when regions with low absorptive capacity cluster together. This obeys the fact that the benefits of the investments directed to accumulate knowledge within the region are not maximized due to the lack of absorptive capacity of the surrounding regions (see e.g. Caragliu and Nijkamp, 2016).

Summing up, we face a setting in which (1) there are (quantifiable) target interactions among firms that enhance the innovation processes, (2) these key interactions could be hindered or neglected by incompatible local industrial policies inherent to the regional divisions, (3) these regional divisions are structural and, in most of the cases, unmodifiable, and (4) it is not sustainable or impossible to relocate firms from one region to another. With this scenario in mind, in this article we propose a quantitative model, the $p$-Innovation Ecosystems ($p$-IE) model, with three main goals in mind: first, to identify the most relevant interactions among existing industries in the country; second, to identify feasible sets of target goods that can improve the economic complexity of the country; and third, to find optimal regional configurations (formed by spatially contiguous administrative units) that maximizes the most relevant interactions among the key industries (according to the previous two elements of the model) within each region. Thus, the $p$-IE model is intended to enable policymakers to apply suitable regional industrial policies that foster innovation processes according to the productive structure of each region. Roughly speaking, in terms of the literature of economic complexity, the $p$-IE model seeks to design regions that improve the local cohesiveness of the regional productive structure according to its position in the PS and, simultaneously, looks for an optimal diffusion branching into the most complex neighboring industries in the PS.

The $p$-IE model is formulated as a mixed-integer programming (MIP) model that belongs to the family of the $p$-regions models devised by Duque et al. (2011). In brief, the $p$-IE involves the aggregation of $n$ areas into $p$ spatially contiguous regions that (1) maximizes the number of links $(l, m)$ in the PS for which industries $l$ and $m$ are within the same region; and (2) identifies strategic relationships between industries within regions that maximizes the probability of activating innovation processes that allow the region to "jump" to more complex neighboring nodes in the PS. Since the $p$-IE model is related to a family of problems that are classified as computationally non-deterministic polynomial-time hard (NP-hard) (Cliff et al., 1975; Keane, 1975), we propose a heuristic solution to effectively compute a near-optimal, if not optimal solution.

From a practical point of view, this approach offers a new way to study the role of regions within countries (e.g., states in United States, departments in Colombia, autonomous communities in Spain, etc.) either as boosters or as obstructions for the key industrial interactions that lead to the improvement of the complexity of the country. More precisely, the existence of regions may affect the evolution of the productive structure in two ways: on the one hand, regional borders between two industries may impair their interaction (as nodes in the PS) diminishing the positive externalities predicted from classical economic theory; on the other hand, the incentives generated by the local administrations could shape the innovation processes of different regions in rather distinct ways enhancing some industrial interactions and discouraging others.

The rest of the paper is presented as follows: Section 2 contains a literature review with the main conceptual elements upon which our model is built up. Section 3 contains a step-by-step construction of the model, the exact formulation, and some illustrative examples. Section 4 presents the heuristic method to solve large instances of this model. Section 5 presents a case of study for Colombia. Finally, Section 6 presents our conclusions and possible future research lines.

# 2 Literature review

The main ideas of this paper are built upon three theoretical bases: (1) complexity theory of Hausmann and Hidalgo, (2) evolutionary economics, and (3) the family of $p$-regions models. This section is basically devoted to discussing the features of these three theories that are relevant for our work.

Hausmann and Hidalgo present in Hausmann and Hidalgo (2011) their theory based upon two main principles: (1) products are the combination of capabilities (of all sorts: inputs, technology, know-how, institutions), and (2) a country can produce a product if and only if it has all the required capabilities to produce the product. Since capabilities are not observable they are determined a posteriori based on two main consequences of their assumptions: (a) the diversity of the products manufactured by a country is related to the diversity and number of capabilities in the country, so that, more capabilities lead to more products; (b) the ubiquity of a product (how many countries produce that product) is a measure of the number of capabilities required by the product.

Although capabilities are quite reminiscent of a general input in the standard production theory in economics, it is important to point out that Hausmann and Hidalgo use them in a more instrumental way. That is, the function that turns capabilities into products is binary (a characteristic function), i.e., it only determines whether the product is produced or not given the availability of the required capabilities within the country (see Hausmann and Hidalgo, 2011). Their approach in Hausmann and Hidalgo, 2011 is also blind with respect to the intensity of the factors, i.e., it only captures the requirement of a factor and not the amount or the way in which the factor is required for production. Hence, their assignation function that relates capabilities with products cannot be thought of as a production function but as a vector of characteristics of each existent good in the global economy (more related to the international trade theory of Heckscher-Ohlin-Vanek, (see Hausmann et al., 2019).

The practical work of Hausmann and Hidalgo is divided into two main blocks, complexity theory and relatedness theory, both underpinned by the concept of capabilities. The concept of economic complexity is derived from a recursive analysis on the duality between diversity (number of exports of a country) and ubiquity (number of countries exporting a product). Diversity per se is not a good measure of the amount of capabilities a country has, since the types of products exported among countries may have substantially different capability requirements depending on the productive characteristics of the exporter. The amount of capabilities required by any product is correlated with the scarcity of the product, i.e., if a product requires a lot of rare capabilities only few countries would be able to produce it. Thus, the interplay of these two concepts, diversity and ubiquity, can determine (as a posterior measure) the amount and the type of existing capabilities within the countries and required by the products. Also, this method helps to determine the degree of complexity embodied in a product. The exact recursive procedure proposed in Hausmann et al. (2014) will be further discussed in Section 3.1.

It is important to note that the economic complexity of a country is determined by the complexity of the products produced by the country. With this in mind, and according to the findings in Hausmann et al. (2014), complex products can be regarded as target products that may be reached (produced) by a country in order to improve its overall complexity and, therefore, its economic growth. The question of how these complex goods can be reached by specific countries is addressed from the relatedness theory that corresponds to the dynamic component of Hausmann and Hidalgo's theory. The dynamic component of Hausmann and Hidalgo's theory appears in Hidalgo et al. (2007) when they seek to understand how countries diversify over time (which, in their terms, is equivalent to accumulating capabilities). They do this with using network theory (the PS): where the nodes are products and the strength of the link is determined by the number of capabilities shared by them, i.e., the link between two products is high if they require a similar set of capabilities to be produced. As such, measuring the relatedness between two goods requires using an observable quantity that captures the capabilities overlapping among goods. However, there is still not a clear consensus on which is the optimal way to measure the relatedness among a set of products. In fact, the study of the structure of relatedness of products within countries and geographic units is still an open field of research in economic geography, mostly in evolutionary economic theory (see Neffke and Henning, 2013; Boschma, 2017).

For the sake of brevity and since it is the only one to be used in our applications; we consider only the PS approach proposed in Hidalgo et al. (2007). Similar constructions of networks measuring the structure of the relatedness of geographic areas can be found in Neffke et al. (2011), OClery et al. (2019), and Diodato et al. (2018). All these approaches are designed so that they measure the co-occurrence (co-production, co-exportation) of goods within the same geographic unit weighted by some quantity (e.g. input requirements, total output) that reflects the extent to which specific economic activities are carried out inside each geographic unit. Explicitly, in Hidalgo et al. (2007) it is proposed a more agnostic or outcome-based approach. Using world trade data, they measure the relatedness using the Revealed Comparative Advantage (RCA) index (Balassa, 1964, see also Section 3.1). Based on this index, the PS is defined as the network of available goods in the world economy with links weighted by the probability of having a comparative advantage among each pair of goods (see Hidalgo et al., 2007). Thus, if the weighted links which define the network are regarded as probabilities to "jump' between the nodes that join them, then the PS not only describes the present output of the country but the actual constraints that face the country to diversify towards other goods in the network. In fact, if we think the position of a country on the PS as the set of products in which it has RCA greater or equal to one, then the innovation dynamics of the countries can be analyzed as diffusion processes on the PS (Hidalgo et al., 2007; Alshamsi et al., 2018; O'Clery et al., 2018]).

Based on their theoretical framework, moving in the PS requires acquiring missing capabilities that combined with the existing ones allows a country to produce a new product and, therefore, move to more complex products in the PS. Thus, the interaction between industries with similar (but not equal) sets of capabilities plays an important role in the acquisition of new capabilities required to produce complex products and move towards better positions in the PS.

Together, the complexity embodied in the products and the relatedness structure of the economic output of the countries provides a tractable framework for longstanding ideas of evolutionary economic theory. Theories like the process of creative destruction proposed by Schumpeter in Schumpeter (1939) can be underpinned quantitatively by the path-dependency inherent to the dynamics on the PS. More explicitly, the evolution led by the natural forces of the economy or the diffusion of the RCA, in terms of the PS, have two main implications: the disappearance of old industries poorly connected in the PS with the local industry and the appearance of new industries naturally associated, according to the PS, with the local industry. This issue regarding the selective evolution of the economic output has been studied empirically in the context of countries (Hidalgo et al., 2007; Hausmann and Hidalgo, 2011; Hausmann et al., 2014), economic regions (Hausmann and Neffke, 2019; Neffke et al., 2011) and cities (OClery et al., 2019; Hausmann et al., 2019) showing that this phenomenon is transversal to any geographic unit in which industry clustering is admissible.

A remarkable empirical work that sheds light on how the process of creative destruction takes place in a geographic (regional) framework is presented in Neffke et al. (2011). In this work they study the evolution of the economic output within 70 industrial regions in Sweden along a period of roughly 30 years (1969-2002). In this work, they measure the evolution of the economic cohesion of the Swedish regions over time. As a measure of cohesion, they use the average of the closeness (in the PS) of each product to the existing portfolio in the region. Intuitively a region is cohesive if there are no incentives for moving, i.e., if the industries within the region are more related among them rather than with industries belonging to other regions. In Neffke et al. (2011) they find that the cohesiveness of the regions has a significant tendency to increase through the years following a process or creative destruction. This work provides a solid confirmation of an idea already hinted by many other studies in complexity theory (Bustos et al., 2012; OClery et al., 2019; Diodato et al., 2018): there is a tendency of the geographical clusters (cities, regions, metropolitan areas) to reproduce the industrial clusters suggested by the PS. In principle, these two types of clusters (geographic and industrial) may disagree because of historical, geographic, cultural or political reasons that encouraged certain types of firms to establish in the region generating an initial endowment of productive activities that shaped the evolution of the economic landscape of the region in the subsequent history. Also, for methodological reasons, it is worth recalling that by construction the PS, at least the one built upon world trade data, captures industrial clusters independently from spatial features of countries and regions.

Another important implication that can be inferred from the works previously discussed is that certain policies that seek to boost local economies by restructuring the economic landscape (such as firm relocation or selective innovation) may be misleading and may have structural issues due to the lack of closeness to the existing portfolio of the target region. In terms of the work of Neffke et al. (2011), for example, the introduction of economic activities weakly linked to existing industries of the region have a higher conditional probability to fail and exit the region. In light of the contributions of complexity theory and evolutionary economics it seems plausible to take advantage of the information encoded in the PS to foster a selective improvement of the productivity (or, more precisely, complexity) as long as the strategy to follow harmonizes with the natural constraints imposed by the topology of the PS. In this order of ideas, and following the arguments presented in the introduction, we opted in this paper for a strategy that involves the design of industrial regions that foster the creation and development of industrial ecosystems. These regions are intended to be spaces in which the interaction among industries is fostered by suitable public policies so that the capabilities required to create new and more complex goods appear naturally according to the dynamics of the PS. This is, the so-called industrial or innovation ecosystems (see also O'Clery et al., 2018).

There is a wide and broad literature about regional planning and region designing (see e.g. Fischer, 1980; Duque et al., 2007); this area has focused in giving objective criteria to define suitable regionalizations that help in the implementation of specific public policies. The main idea of this approach consists of decomposing global policy objectives that are applied into big geographic units into local tasks in specific regions which integrates in a harmonic way and finally promotes the achievement of the policy goal (Fischer, 1980).

Among the formulations for this problem are linear optimization models (Garfinkel and Nemhauser, 1970; Zoltners and Sinha, 1983), mixed-integer programming (Duque et al., 2011) and nonlinear models (Macmillan and Pierce, 1994), and the use of techniques such as implicit enumeration of feasible solutions (Garfinkel and Nemhauser, 1970) and column generation (Mehrotra et al., 1998). In particular, our model follows the lines of Duque et al. (2011) which is formulated a mixed integer programming (MIP) problem, called the $p$-regions, for aggregating $n$ areas into $p$ spatially contiguous regions, using three different strategies to ensure continuity. One of the main advantages of the $p$-regions type models is their flexibility to incorporate spatial contiguity without adding constraints on the shape of the resulting regions. This, in turn, allows them to capture a diverse range of spatial patter (compact, elongated, among others). For example, based on the $p$-regions model there have been different MIP formulation by adding additional features, such as maximizing the number of regions being added (Duque et al., 2012), generating $p$ regions while maximizing their compactness (Li et al., 2014), defining $p$-functional regions (Kim et al., 2015), among others. Such formulations ensure that we can obtain an optimal solution. However, they are computationally intensive to solve and sometimes it is better to use heuristic approaches for resolving large problem instances. A good heuristic should be capable of solving a problem of region delineation, within a shorter amount of computation time, with the goal of identifying feasible solutions as close as possible to the optimum (Duque et al., 2007).

In this work, we take advantage of the versatility and robustness of the economic complexity theory of Hausmann and Hidalgo and embed it into a region designing model. Thus, we propose to draw a conceptually and theoretical link between these two subjects providing a new model with a solid economic foundation to the $p$-regions family, based on Hausmann and Hidalgo's theory and evolutionary economics. This also allows us to incorporate new elements into Hausmann and Hidalgo's theory such as the role of the regions in the innovation process followed by the countries on the PS. Also, a proved relationship between the regional division of a country and its economic growth will open the possibilities for developing new economic-based models to allow policymakers to: (1) identify how far is the actual regional configuration of its economy from the optimal one that would potentially foster economic growth, and (2) design industrial regions, or intraregional agreements between administrative units, to minimize the boundary effects and ease the interaction among the key regions to promote the generation of innovation ecosystems.

# 3 The model

In this section we present the elements and the step-by-step assembling of the objective function of the $p$-IE model. For the sake of clarity, we divide this section in 4 parts. The first part provides an explanation of the inputs of the model: elements of the PS (economic) and spatial elements (geography); also, it shows how these elements are systematically incorporated into the objective function. The second part provides a summarized statement of the optimization problem. Part three contains the final expression that will be maximized. Finally, in part four we illustrate the properties of the solution of the p-IE model by means of some suitable toy examples.

## 3.1 Conceptual framework

Complexity theory, in this work, plays the role of giving a quantitative framework to the process of identifying key industries. There are two types of key industries: (1) the ones that already exist and whose interaction is central in the given economy, and (2) complex industries that are yet to appear in the economy of interest. Our measurement tool in this case is the Product Space (PS), which describes the relatedness structure between all the goods in the world (or in another macrogeographic reference, e.g. a continent, a set of countries, etc.) and therefore provides information about classes of key industries.

Mathematically, we will view the PS together with the complexity theory as a weighted network of products, where the weight in the nodes gives a measure of the relevance of each good (1 the most relevant and 0 the least relevant) and the weight in the edges gives a normalized notion of association between any pair of goods. It is important to stress the fact that even if we consider a weighted network built upon Hausmann and Hidalgo's theory (Hausmann et al., 2014; Hidalgo et al., 2007), any network with the properties specified above would work for our purposes.

#### 3.1.1 First term: relatedness structure and key industrial links

The PS proposed in (Hausmann et al., 2014) measures the relatedness by means of the Revealed Comparative Advantage (RCA) index introduced by Balassa (Balassa, 1964) which is given by

$$\mathbf{RCA}_{c,p} = \frac{x_{cp}/\sum_p x_{cp}}{\sum_c x_{cp}/\sum_{c,p} x_{cp}}$$

Where $x_{cp}$ is the amount of the good $p$ exported by the country $c$ and the sums runs over all the possible goods a countries. Given these indices for each good and each country, Hausmann et al. (2014) defines the degree of relatedness between goods $k$ and $l$ as:

$$y_{lm} = \min\{\mathbb{P}(\mathbf{RCA}_{m,p} > 1 | \mathbf{RCA}_{l,p} > 1), \mathbb{P}(\mathbf{RCA}_{l,p} > 1 | \mathbf{RCA}_{m,p} > 1)\} \tag{1}$$

That is, the minimum of the conditional probability of obtaining RCA in a good given that the country already has RCA in the other good (the minimum is taken so that the weight is undirected).

Hence, the PS is defined as the undirected network (see Figure 1a) of available goods in the world economy with links weighted by the numbers $y_{lm}$ whenever these weights exceed a certain threshold (see Hausmann et al., 2014, or Section 5 for a brief discussion about the determination of the threshold).

Following the ideas of Hausmann et al. (2014) we can assess the overall production of a country from its position on the PS. The idea consists in determining the set of goods in which the country has revealed a comparative advantage, i.e., RCA greater than 1. We regard this set as the initial position of the country on the graph (on the PS) and we will denote it by $IG$ (Initial goods), analogously we denote by $IL$ the set of initial links induced in the PS by $IG$, i.e., the links of the PS connecting nodes belonging to $IG$ (see Figure 1b).

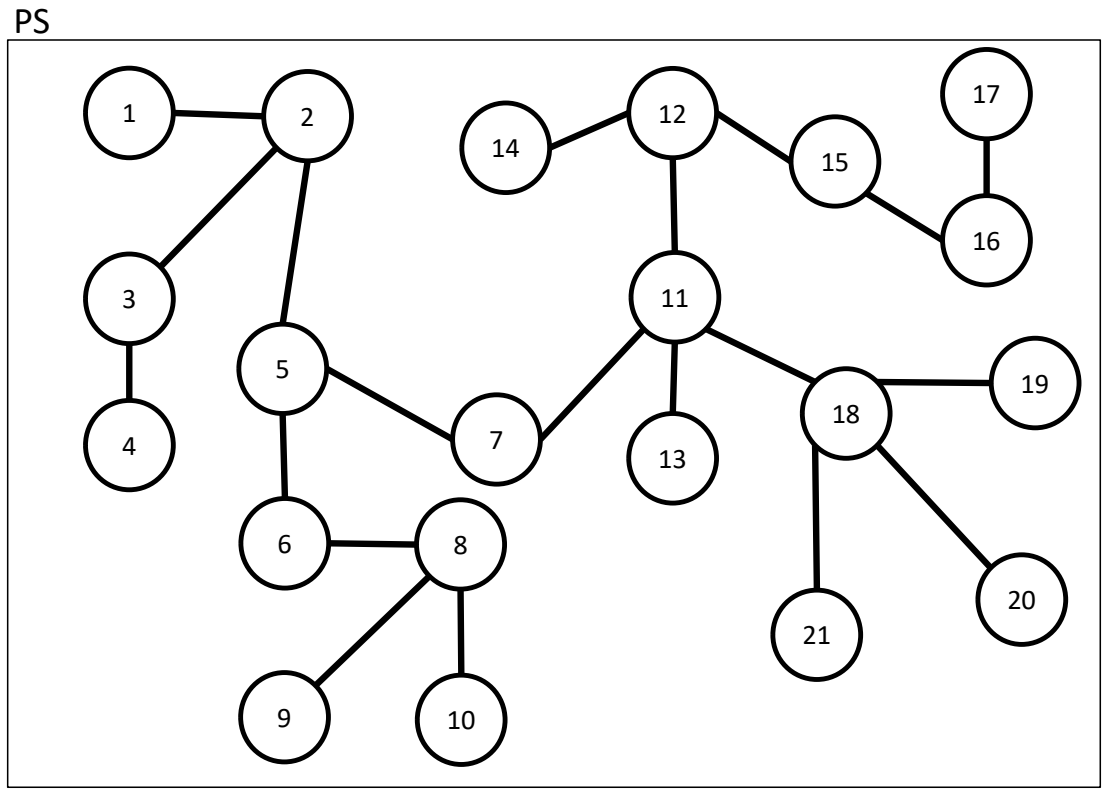

(**a**) Layout of the PS of an abstract economy with 21 goods.

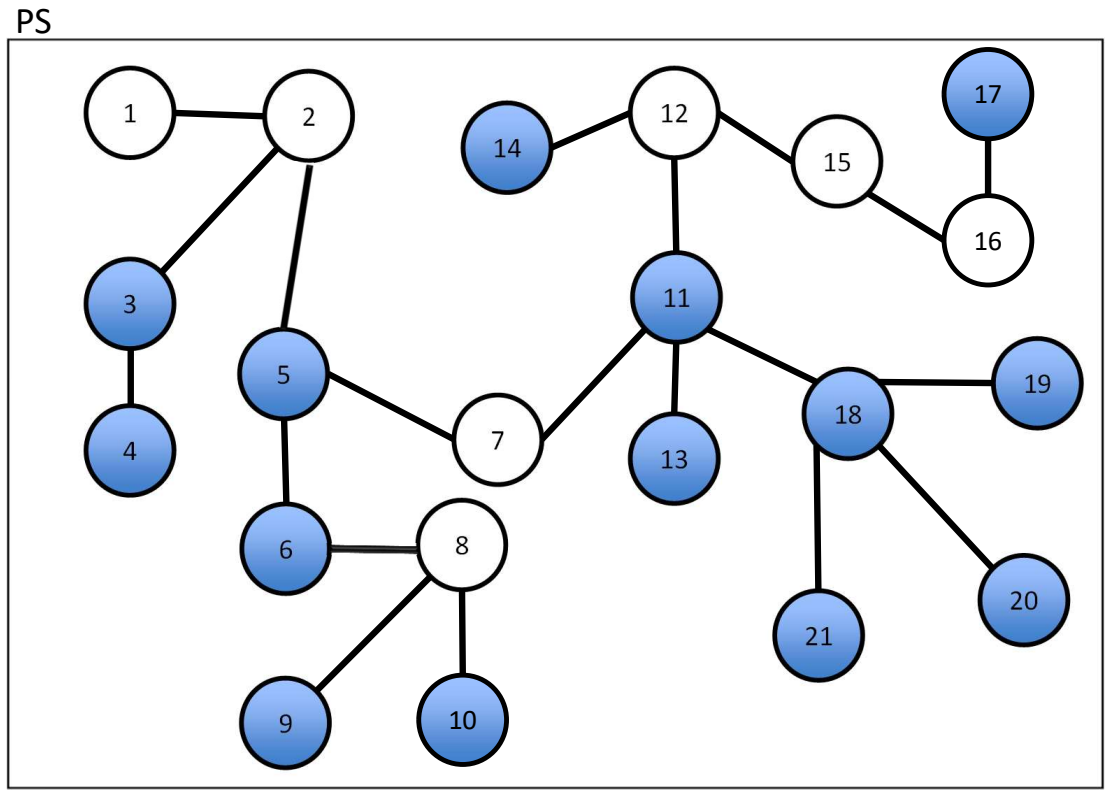

(**b**) Available (blue) goods in the economy $IG$ determine the position of the country on the PS.

**Figure 1:** The first part shows the PS faced by the economy, whilst the second part shows in blue the goods in which the economy has comparative advantage.

Recalling the ideas discussed in the literature review about the role of regions in the formation of clusters and on the diffusion on the PS, we introduce geography in this model by means of two effects: boundary effect and distance decay. Since production and, subsequently, spillover effects measured by the Hausmann and Hidalgo's theory takes place in very concrete spatial units, we incorporate the boundary effect by measuring the interaction between industries which have an active presence within the same region. These measures are derived from two spatial inputs: the division of the country (or the general spatial framework of reference) into areas (see Figure 2a) and the frequency matrix $f = (f_{li})$ of the industries among the areas (see Figure 2b) (value of exports in this case, since we follow Hausmann et al., 2014).

| | | | | |
|---|---|---|---|---|
| 1 | 2 | 3 | 4 | 5 |
| 6 | 7 | 8 | 9 | 10 |
| 11 | 12 | 13 | 14 | 15 |
| 16 | 17 | 18 | 19 | 20 |

(**a**) Country with 20 areas.

| Area/Industry (j,l) | 3 | 4 | 5 | 6 | 9 | 10 | 11 | 13 | 14 | 17 | 18 | 19 | 20 | 21 | 22 |
|---|---|---|---|---|---|---|---|---|---|---|---|---|---|---|---|
| 1 | 100 | | | | | | | | | 100 | | | | | |
| 2 | | | | 100 | | | | | | | | | | | |
| 3 | | | | | | | | | | | 100 | | 100 | | |
| 4 | | | | | | | | | | | | | 100 | 100 | |
| 5 | | | | | | | | | | | 100 | | 100 | | |
| 6 | | 100 | | | | | | | | | | | | | |
| 7 | | | 100 | | | | | | | | | | | | |
| 8 | | | | | | | | | | | 100 | | | | 100 |
| 9 | | | | | | | | | | | 100 | | | | |
| 10 | | | | | | | | | | 100 | | 100 | | 100 | |
| 11 | 100 | | | | | | | | | | | | | | |
| 12 | | | | 100 | | | | | | | | | | | |
| 13 | | | | | | | | 100 | | | | | | | |
| 14 | | | | | | | 100 | | | | | | | | |
| 15 | | | 100 | 100 | | | | | | | | | | | |
| 16 | | 100 | | | | | | | | | | | | | |
| 17 | | | 100 | | | | | | | | | | | | |
| 18 | | | | | | | 100 | | 100 | | | | | | 100 |
| 19 | | | | | | 100 | | | 100 | | | | | | |
| 20 | | | | 100 | 100 | | | | | | | | | | |

(**b**) Distribution matrix $f = (f_{lj})$ for an economy with 22 industries ($l = 1, \cdots, 22$) and 20 areas ($j = 1, \cdots, 20$).

**Figure 2:** The first part depicts the geographic distribution of the 20 areas in which each area is given by a square of side length 1. The second part shows the distribution of 22 industries in the country. Notice that we omit the columns corresponding to the industries 1 and 2 since the country do not produce any of them.

From the geographic distribution of the firms we compute the relative relevance of area i producing good l (that is the share of an area in the total production, computed from Figure 2b)

$$r_{il} = \frac{f_{il}}{\sum_i f_{il}} \tag{2}$$

We also incorporate a distance decay factor that penalizes the interaction of industries located far away one from the other. This geographic effect is justified theoretically in Hausmann and Hidalgo (2011) by the strong effect of distance in the, arguably, more important capability: knowledge. We propose a decay function $f : [0, \infty) \mapsto [-1, 1]$ of the form

$$f(x) = 1 - 2\left(\frac{x}{M}\right)^{\alpha}, \tag{3}$$

where $M$ is the maximal distance between two areas in the country and $\alpha$ is the distance decay parameter. Notice that $f(0) = 1$ and that $f(M) = -1$. The parameter $\alpha$ determines how fast the decay function becomes negative, decaying faster when $\alpha$ is greater than 1 and slower whenever $\alpha$ approaches to 0 (see Section 3.4 and Section 5 for general rules to establish the value of $\alpha$). Thus, we capture the interaction between the goods $l$ and $m$ produced in the areas $i$ and $j$, respectively, using the following expression

$$t_{ij}^{lm,k} r_{li} r_{mj} f(d_{ij}), \tag{4}$$

where $d_{ij}$ is the distance between area $i$ and $j$, $t_{ij}^{lm,k}$ is an indicator function that is 1 whenever $i$ and $j$ produce goods $l$ and $m$, respectively, and both belong to the same region $k$ (see Figure 3). Hence, in a setting with $n$ areas and $p < n$ regions, the first term of our objective function would have the form

$$FT(S_p) = \sum_{k=1}^{p} \sum_{i=1}^{n} \sum_{j=i}^{n} \sum_{(l,m)\in LI} t_{ij}^{lm,k} r_{li} r_{mj} f(d_{ij}) \tag{5}$$

where $S_p$ is any partition of the $n$ areas into $p$ regions which in turn determine the value of the variables $t_{ij}^{lm,k}$.

We stress the relevance of two features in the structure of the first term of the objective. First, notice that we require the inverse function of the distance $f$ to assign negative values to penalize long distances with negative values (see Section 3.4 for a further discussion). Second, we capture the interaction among areas in a multiplicative way to neglect all the cases in which at least one of the intensities is 0. This idea and the role of the indicator functions $t_{ij}^{lm,k}$ are fully explained in Figure 3. We describe five possible scenarios of interaction: the first one, in which both, areas $i$ and $j$, produce two industries linked in the PS ($l$ and $m$), yielding 2 self-interactions within each area ($r_{li} r_{mi}$ and $r_{lj} r_{mj}$) and two unidirectional interactions between industries $l$ and $m$ across areas ($r_{li} r_{mj}$ and $r_{mi} r_{lj}$); the second scenario exhibits one loop interaction within area i between industries $l$ and $m$ ($r_{li} r_{mi}$) and the interaction between industry $l$ in area $j$ and industry $m$ in area $i$ ($r_{mi} r_{lj}$); the third scenario presents one unidirectional interaction; scenarios four and five show that neither the presence of only one industry in one region nor the presence of the same industry in several areas do not yield any interaction, respectively.

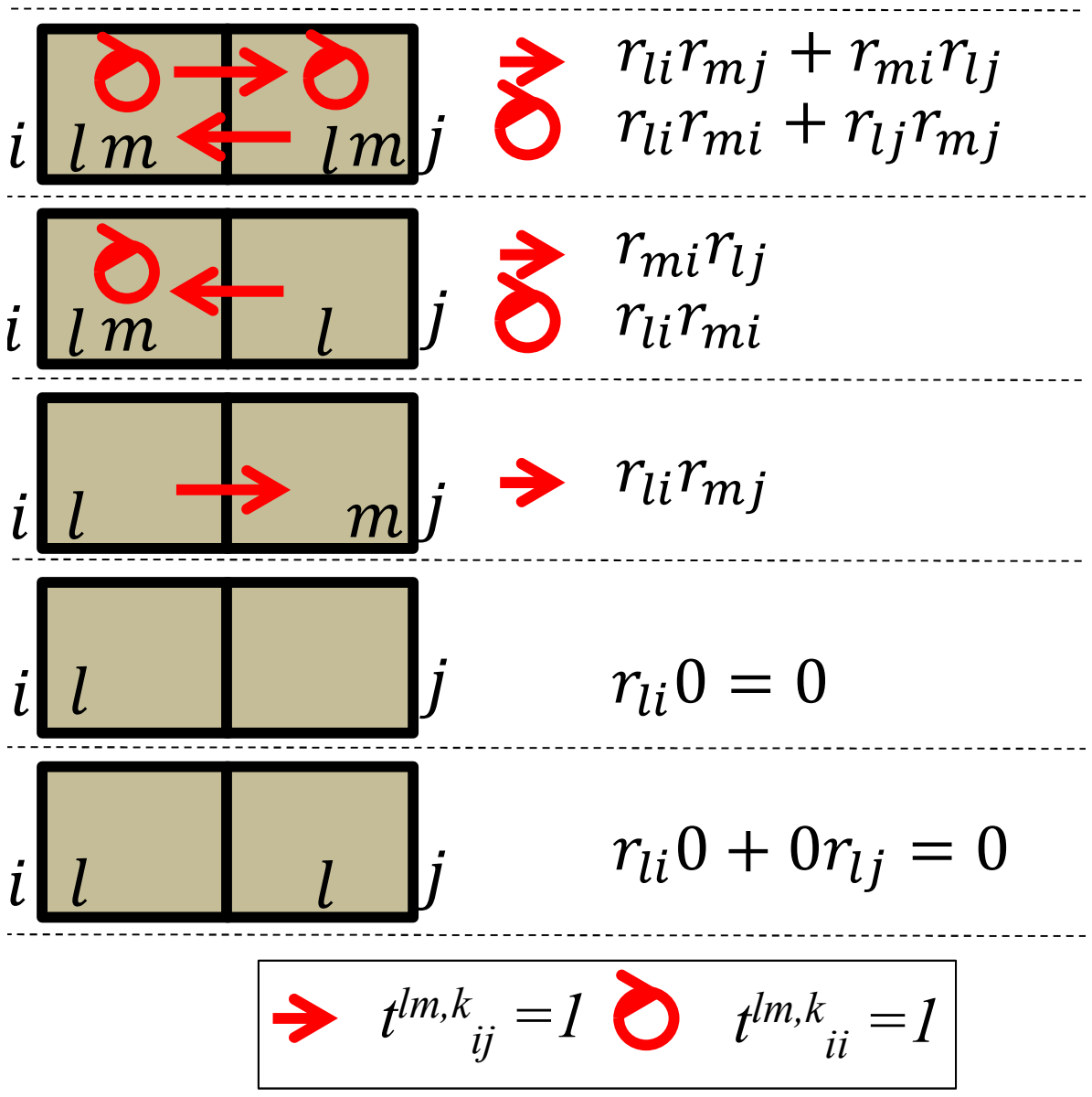

**Figure 3:** Fixing a region $k$, the indicator functions $t_{ij}^{lm,k}$ capture the self-interactions within areas (loops) and the interaction between pairs of areas $(i, j)$ ) (arrows) producing linked goods $(l, m)$ in the PS.

### 3.1.2 Second term: complexity and target products

While the first component enhances the interaction among the goods that already exist in the economy, the second component is designed to maximize the probability to reach new products in the PS that are not yet produced by the country. The construction of this second part relies in two basic principles: (1) nodes with a bigger weight (more complex) are more desirable for the economy and (2) the pairwise interaction of products with high association to a third non-existent good in the economy increases the probability of the arising of this new good, whenever the interaction is given within the same region. More generally, the problem addressed by the second term of the objective functions lies in the general framework of a directed complex contagion problem, i.e., a diffusion process on the PS in which the process can diffuse from several nodes to others depending on their complexity. Such models predict a nonlinear diffusion, more explicitly, the probability of reaching a new node turns out to be proportional to a power of the association with the first order neighbors (see Alshamsi et al., 2018). Nonetheless, in Alshamsi et al. (2018) the contagion power for the PS was found to be conveniently close to 1 (1.03, more precisely), enabling us to consider the sum of pairwise interactions. Notice also that the study of contagion as a pairwise interaction of nodes is quite reminiscent to the basic concept of clustering by triplets on networks (see e.g. the definition of the clustering coefficient in Barabási et al., 2016). For our subsequent analysis we introduce the notation $L_v$ for the set of (undirected links) of neighbors of $v$ in the PS. Notice that two nodes linked in $L_v$ may not (but usually are) linked in the PS.

Since the non-existent goods may be far away from the set $IG$ in the PS, we restrict our analysis to products that can be reached after one step of diffusion, trying to simplify the study of more complex dynamics. Also, since this component of the objective function is intended to improve the complexity, then we define the target goods as the ones that are more complex than at least one of their neighbors in the PS already produced by the country. Further, the spatial nature of this model allows us to only consider complex goods whose arising requires the interaction of two or more industries, i.e., we discard leaves of the network. We depict thoroughly these processes in Figure 4a. This set of nodes in the graph constitutes the set of target goods denoted by $TG$. Although many complexity measures would work for our purposes, we want to stress that its introduction in this model is motivated by the definition of the Economic Complexity Index (ECI) introduced in Hausmann et al. (2014). In our case of study (see Section 5), we use a normalized version in the unit interval of this index as a measure of complexity.

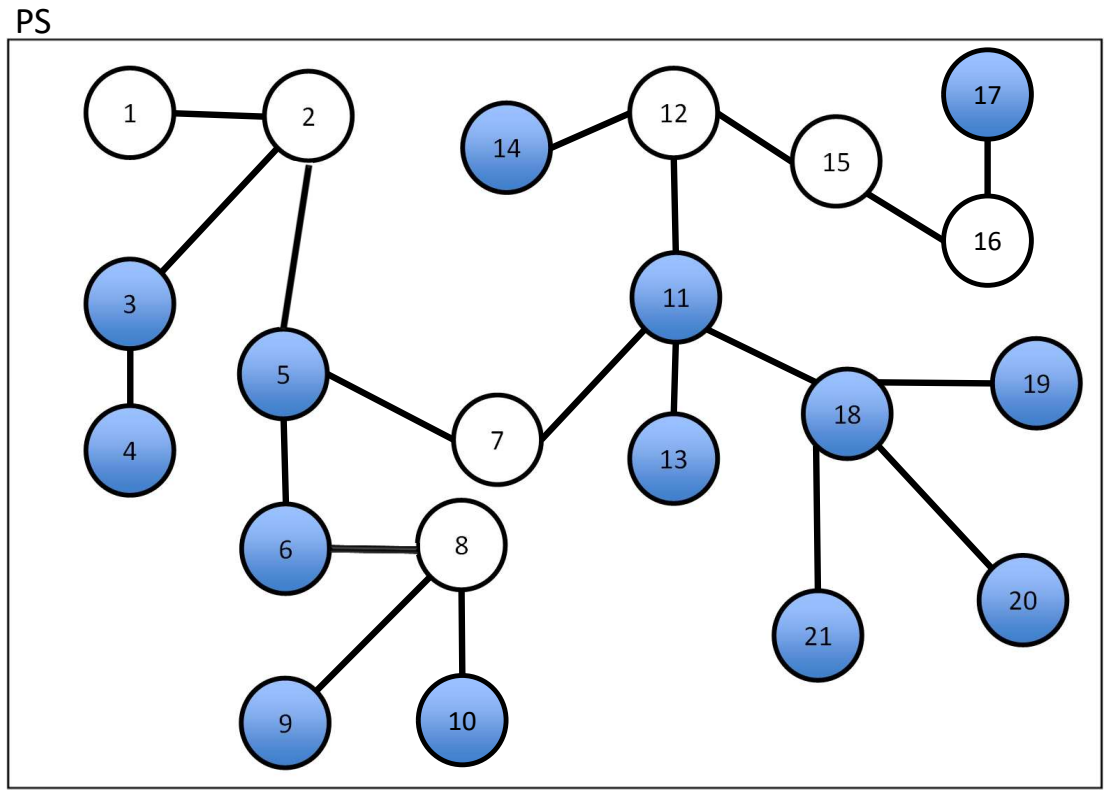

**(a)** First order neighbors of $IG$.

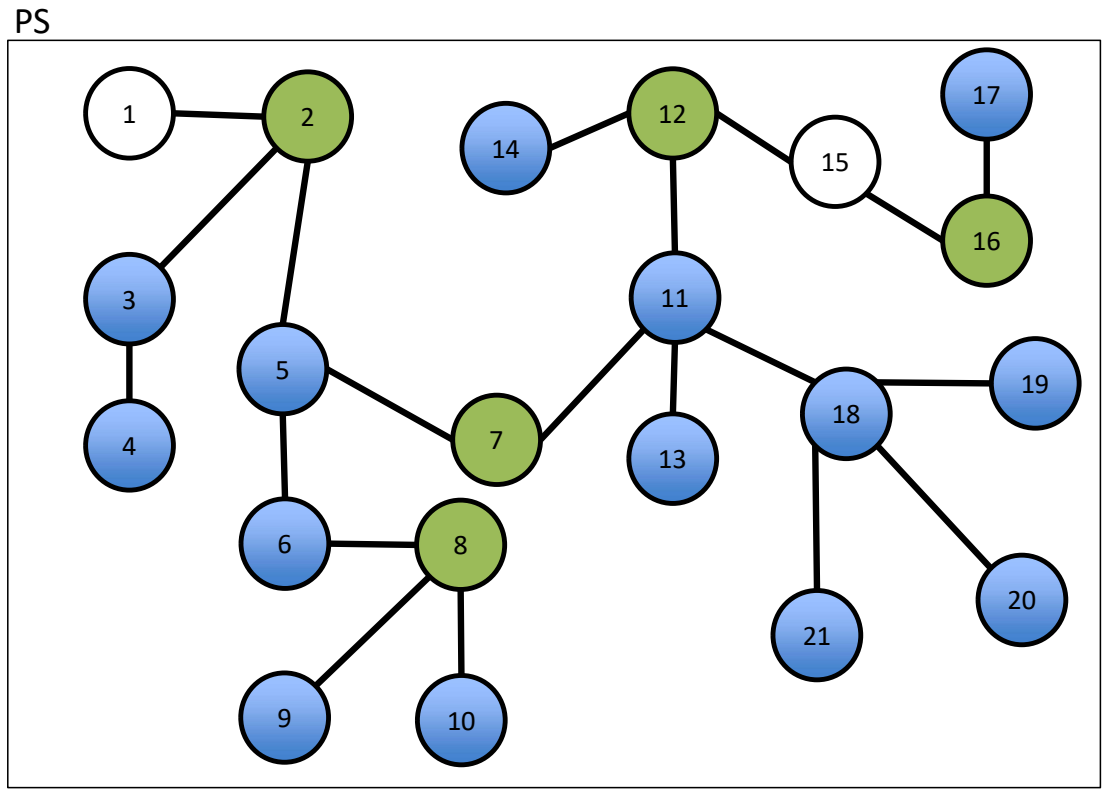

**(b)** Selection of the target goods $TG$.

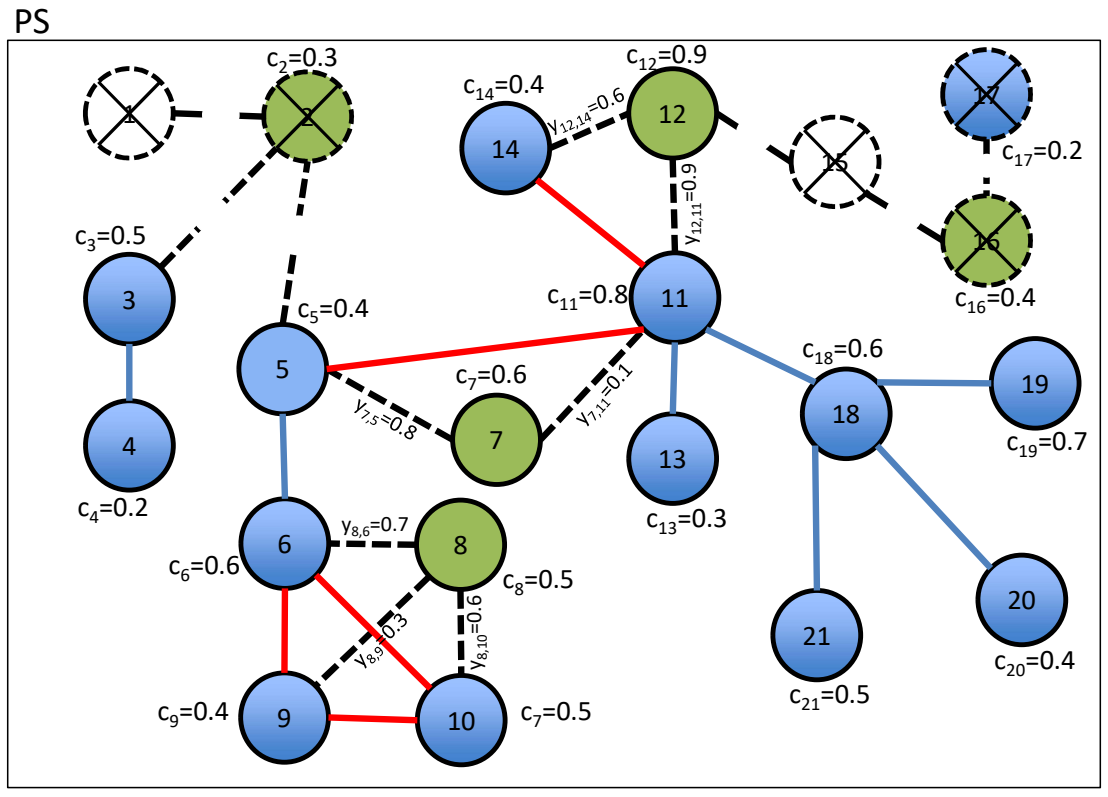

**(c)** Final input network.

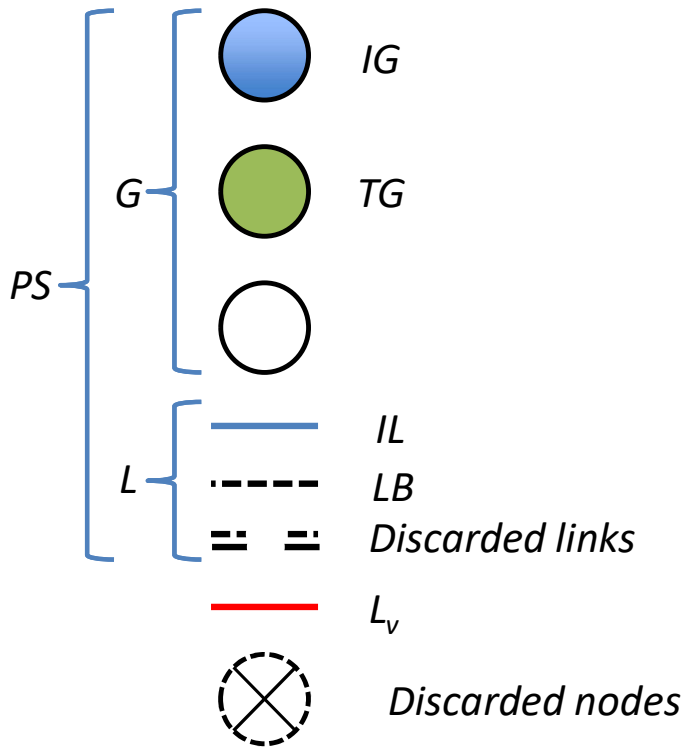

**(d)** Nomenclature.

**Figure 4:** The first three parts depict how the target nodes ($TG$) are chosen. In Figure 4a we select the first order neighbors of the blue nodes $IG$. In Figure 4b we discard the higher order neighbors (nodes 1 and 15), the leaves of the network that do not require spatial interaction to be produced (node 16), and the green nodes that do not improve the complexity of the system (node 2). Figure 4c depicts the final version of the PS and Figure 4d summarizes the nomenclature used in for this selection process.

Thus, arguing as in the formulation of the first part we have that the second component should have the form

$$ST(S_p) = \sum_{k=1}^{p} \sum_{v \in TG} \sum_{i=1}^{n} \sum_{j=1}^{n} \sum_{(l,m) \in L_v} t_{ij}^{lm,k} (y_{lv} r_{li} + y_{mv} r_{mj}) f(d_{ij}) c_v \tag{6}$$

where $c_v$ is the complexity of the good $v \in TG$, $y_{lv}$ is the weight of the link between the goods $l$ and $v$ in the PS (equation (1)).

### 3.1.3 Assembling the objective function

So far, we have set up both components of our objective function without specifying the interplay between them. By construction, the first term enables us to capitalize the key industrial interactions already formed by the natural economic forces of the country of interest. Since the second term tries to reach as many complex goods as possible it has the tendency to group multiple areas and create big regions while introducing volatility into the model. Thus, the first term plays the role of guaranteeing a reasonable size and shape of the economic regions so that key interactions are preserved, whereas the second term is intended to "perturb" or modify the configurations suggested by the first term in order to maximize the complexity without making abrupt changes in regional configurations. Hence, the next step consists in finding suitable weights in order to guarantee the predominance of the first term over the second one.

Notice that the terms in equations (5) and (6) are bounded from above by 2. Hence, in order to bound this term, it is necessary to determine which is the maximum number of ways to assign $n$ areas into $p$ regions.

This problem can be phrased in terms of a maximization problem. Let $z_i$ be the number of areas in the region $i$. From Figure 3 we have that each area can interact with all the areas, including itself. So, if we count the number of interactions we have that the first area can interact with $z_i$ areas, the second with $z_i - 1$ and so forth. Thus, we face the constrained optimization problem

$$\begin{aligned} \text{Maximize } & \frac{1}{2} \sum_{i=1}^{p} z_i (Z_i - 1), \\ \text{s.t. } & \sum_{i=1}^{p} z_i = p, \\ & z_i \geq 1. \end{aligned}$$

In here, the objective function measures the number of interactions between all areas in each region, i.e., the number of $t_{ij}^{lm,k}$ that can be 1 assuming that all goods are produced in all the areas and that two areas can be joined in only one region (where this latter restriction will be guaranteed by the constraints of the model). A direct computation shows us that the solution of this maximization problem $\frac{n}{2p(n-p)}$. Thus, the first component of the objective function is bounded from above by $\frac{n}{2p(n-p)}|IL|$, where $|IL|$ is the number of elements in $IL$. A similar reasoning applied to the second term shows that this term is bounded from above by $h = \frac{n}{2p}(n-p)\sum_{v \in TG} |L_v|(|L_v|-1)$.

Hence, a possible factor that guarantees the dominance of the first term of the objective function over the second term would be $W = 10^{1+\lfloor \log(h) \rfloor}$ (where $\lfloor \cdot \rfloor$ is the standard floor function.) We can also normalize the weights setting $W_1 = \frac{W}{W+1}$ and $W_2 = 1 - W_1$. Finally, given a $p$-regionalization $S_p$ (or partition of $n$ areas into $p$ regions) we define the objective function of the $p$-IE by

$$Z(S_p) = W_1 FT(S_p) + W_2 ST(S_p) \tag{7}$$

## 3.2 Problem statement

Let $I = \{1, \cdots, n\}$ be the set of areas of a country with $n$ areas and let $p \in \{2, \cdots, n-1\}$ be the prescribed number of regions. Let $\Pi$ be the set of feasible $p$-regionalizations given by contiguous partitions of $I$, i.e., collections of areas the form $P_p = \{R_1, \cdots, R_p\}$ such that

- $R \neq \emptyset$ for $R \in P_p$
- $R \cap R' = \emptyset$ for any pair of distinct elements $R, R' \in P_p$.
- $\bigcup_{R \in P_p} R = I$.
- Each region $R \in P_p$ is connected.

The $p$-IE problem may be formulated as:

$$\text{Determine } P_p^* \in \Pi \text{ such that: } Z(P_p^*) \geq Z(P_p), \ \forall P_p \in \Pi.$$

## 3.3 The exact formulation of the $p$-Innovation Ecosystems problem

Now we are in position to formulate the exact model.

**Parameters**:

** The geography*:

| | |
|---|---|
| $A =$ | set of areas, $A = \{1, \cdots, n\}$ ; |
| $i, j =$ | the indices used to refer to specific areas, where $i, j \in A$; |
| $n_{ij} =$ | $\begin{cases} 1, \text{if areas } i \text{ and } j \text{ share a border, with } i \neq j; \\ 0, \text{otherwise;} \end{cases}$ |
| $N_i =$ | $\{j \mid n_{ij} = 1\}$ , the set of areas that are adjacent to area $i$; |
| $d_{ij} =$ | the distance between areas $i$ and $j$, where $i < j$; |
| $f =$ | a decreasing function in the distance between areas $i$ and $j$ with range $[-1, 1]$; |
| $k, K =$ | the index and set of regions, $K = \{1, \cdots, p\}$ ; |
| $o =$ | the index used to refer to the contiguity order, $o \in \{0, \cdots, q\}$ , with $q := n - p$. |

* *The Product Space*:

| | |
|---|---|
| $PS =$ | any instance of the Product Space (PS); |
| $g, G =$ | the index and set of goods/nodes in $PS$; |
| $l, m, v =$ | the indices used to refer to specific goods, where $l$, $m$, $v \in G$; |
| $c_v =$ | the complexity of good $v$ ; |
| $L, \{l, m\} =$ | the set $L$ in which each element $\{l, m\}$ indicates a link connecting goods $l \in G$ and $m \in G$ in $PS$; |
| $y_{lm} =$ | the weight in the PS between the goods $l$ and $m$; |
| $IG =$ | set of available goods for the country under study, which are connected in the PS given certain threshold in the weights of the links; |

$TG =$ set of goods/nodes $g \in G$ that the country under study defines as target goods. $GT \subseteq G$ (see equation 6). The set $TG$ satisfies the following properties:
$TG \cap IG = \varnothing$;
$TG \cup IG \subseteq G$;
each good in $TG$ is a first order neighbor of, at least, two goods in $IG$;
i.e., for each good $l \in TG$ there are links $l, m$ in $PS$ such that $m \in IG$;
each good $v$ in $TG$, has a $c_v$ that is greater than the minimal complexity of their neighboring goods in $IG$ (there is an improvement in the complexity);

$IL =$ the set of links $\{l, m\}$, such that $l \in IG$ and $m \in IG$. $IL \subseteq L$;

$LB =$ the set of links $\{l, m\}$, such that $l \in IG$ and $m \in TG$. $LB \subseteq L$;

$L_v =$ the set of links $(l, m)$ with the pairwise connections between the goods $l, m \in IG$ such that both are connected in the PS to $v \in TG$. These links are not of the same nature as the links in the PS.

* *The relationship between the geography and the Product Space*:

$r_{li} =$ The relative relevance of the area $i$ producing the good $l$, see equation (2).

$$b_{li} = \begin{cases} 1, \text{ if } r_{l,i} > 0 \\ 0, \text{ otherwise.} \end{cases}$$

* *Instrumental parameters*:

$h =$ $h = \frac{n}{2p}(n-p) \sum_{v \in TG} |L_v|(|L_v| - 1)$,

$W =$ $10^{1+\lfloor \log(h) \rfloor}$,

$W_1 =$ $\frac{1+W}{W}$,

$W_2 =$ $1 - W_1$.

Decision variables:

$$t_{ij}^{lm,k} = \begin{cases} 1, \text{if areas } i \text{ and } j \text{ belong to region } k, \text{area } i \text{ produces good } l \\ \quad \text{and area } j \text{ produces good } m \\ 0, \text{otherwise.} \end{cases}$$

$$x_i^{ko} = \begin{cases} 1, \text{if areas } i \text{ is assigned to region } k \text{ in order } o \\ 0, \text{otherwise.} \end{cases}$$

Maximize:

$$\begin{aligned} Z =& W_1 \sum_{k=1}^{p} \sum_{i=1}^{n} \sum_{j=i}^{n} \sum_{\{l,m\}\in LI} t_{ij}^{lm,k} \left( r_{li} r_{mj} y_{lm} \right) f\left( d_{ij} \right) + \\ & W_2 \sum_{k=1}^{p} \sum_{v\in GT} \sum_{i=1}^{n} \sum_{j=i}^{n} \sum_{\{l,m\}\in L(L_v)} t_{ij}^{lm,k} \left( y_{lv} r_{li} + y_{mv} r_{mj} \right) c_v f\left( d_{ij} \right). \end{aligned} \tag{8}$$

Subject to:

$$\sum_{i=1}^{n} x_i^{k0} = 1 \qquad \forall k = 1, \cdots, p, \tag{9}$$

$$\sum_{k=1}^{p} \sum_{o=0}^{q} x_i^{ko} = 1 \qquad \forall i = 1, \cdots, n, \tag{10}$$

$$x_i^{ko} \leq \sum_{j\in N_i} x_j^{k(o-1)} \qquad \forall i = 1, \cdots, n; \forall k = 1, \cdots, p; \forall o = 1, \cdots, q, \tag{11}$$

$$2t_{ij}^{lm,k} \leq \sum_{o=0}^{q} x_i^{ko} b_{li} + \sum_{o=0}^{q} x_j^{ko} b_{mj} \qquad \forall i, j = 1, \cdots, n; \forall k = 1, \cdots, p; \forall \{l, m\} \in LI \cup L_v, \tag{12}$$

$$t_{ij}^{lm,k} \geq \sum_{o=0}^{q} x_i^{ko} b_{li} + \sum_{o=0}^{q} x_j^{ko} b_{mj} - 1 \qquad \forall i, j = 1, \cdots, n; \forall k = 1, \cdots, p; \forall \{l, m\} \in LI \cup L_v; f(d_{ij}) < 0, \tag{13}$$

$$x_i^{ko} \in \{0, 1\} \qquad \forall i = 1, \cdots, n; \forall k = 1, \cdots, p; \forall o = 0, \cdots, q, \tag{14}$$

$$t_{ij}^{lm,k} \in \{0, 1\} \qquad \forall i, j = 1, \cdots, n; \forall k = 1, \cdots, p; \forall \{l, m\} \in LI \cup L_v. \tag{15}$$

Constraints (9), (10) and (11) guarantee the connectedness of each defined region. These conditions represent an extension of the ordered area assignment conditions proposed by Cova and Church (2000) ( see also Martin and Sunley, 1996). More precisely, equation (9) forces the model to assign only one area per region with order 0, equation (10) implies that each area must be assigned with some order to some region, and equation (11) preserves the order of the assignment of the areas to the regions, i.e., guarantees that an area is assigned to a region with order o only if there is another neighboring area belonging to the same region assigned with order o-1.

Restrictions (12) and (13) provide the interaction between the regionalization ($x_i^{ko}$) and the interaction in the product space ($t_{ij}^{lm,k}$ ). Equation (13) bounds the decision variables $t_{ij}^{lm,k}$ so that the model is allowed to activate them (and increase the value of the OF) when the interaction between the goods $l$ and $m$ produced in the areas $i$ and $j$ (respectively) occurs within the same region $k \in \{1, \cdots, p\}$ , i.e., this restriction generates the trade-off in the model that guarantees that only the most important links of the product space are preserved. On the other hand, we use equation (13) to force the model to activate all the variables $t_{ij}^{lm,k}$ whenever areas $i$ and $j$ are within the same region and their interaction is penalized by the distance decay function $f$ so that the OF has to pay effectively for the generation of large regions. This restriction prevents the model in producing degenerate solutions, i.e., one massive region and $p - 1$ small regions. Controlling the size of the regions through the distance decay allows us to formulate the model for any geography without the risk of having empty feasible sets (infeasibilities). Finally, equations (14) and (15) are integer conditions on the decision variables.

## 3.4 Examples

We present two examples to illustrate how our model works. In the first example we show the behavior of the first term of the OF, whereas the second shows the influence of the second term on the solution: explicitly, how the second term of the OF breaks ties in a solution reached by the first term, seeking to maximize complexity. In both cases, we will take as distances between areas the Euclidean distance between the centers of the regions.

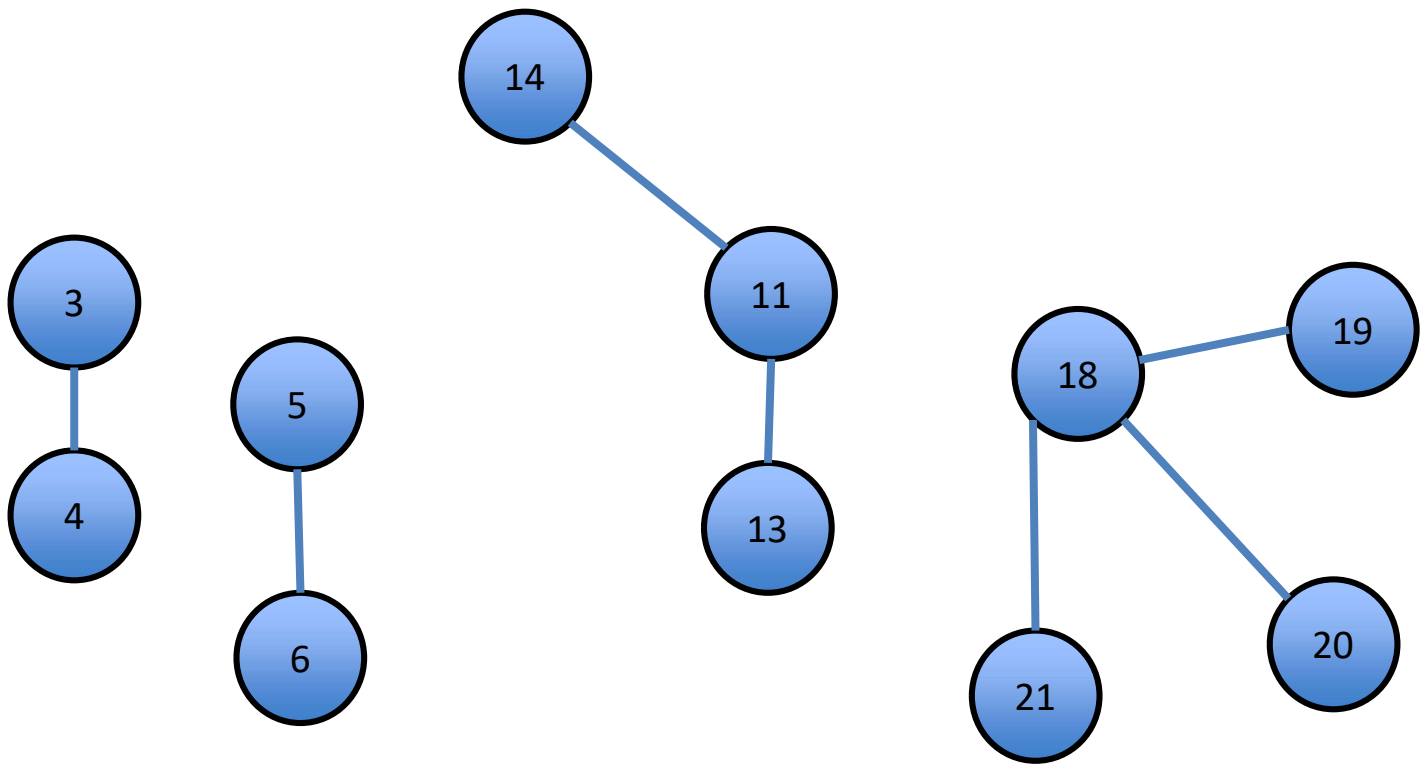

**Figure 5:** PS for example 1. This is the PS of the country described in Figure 2. It contains 4 clusters of industries perfectly separated and omits industries 9, 10 and 17 which are isolated.

We will take $\alpha$, i.e., the decaying parameter of the function $f$ in equation (3), such that it gives zero value to the radius of the region in the scenario in which all of them have the same number of areas $\frac{n}{p}$. Thus, since the geography in these toy examples is given by square areas in rectangular grids (see Figure 6b and Figure 7c) we will take $R_{av} = \sqrt{\frac{n}{p}}$ as the radius of an average region. Therefore, we have that $\alpha = \frac{ln(2)}{ln\left(\frac{M}{R_{av}}\right)}$.

For the first example we suppose a country with geography and production described in Figure 2 and, therefore, we have the decaying parameter $\alpha = 0.7564$. Consider also the PS depicted in Figure 5. Notice that the PS given in Figure 5 is a subset of the PS presented in the Figure 1, although some links are removed or added so that we obtained 4 disconnected clusters on the graph. Also notice that in this case the graph is entirely exhausted by the set $IG$, equivalently, there are no target goods. Hence, the weights of the vertices are irrelevant for the objective function. From the conveniently designed distribution of industries on the array, there are five spatial clusters of industries where each one of those corresponds to one of the four connected components of the PS in the Figure 5. Thus, for p=5 the $p$-IE model finds an optimal solution of the form presented in Figure 6a.

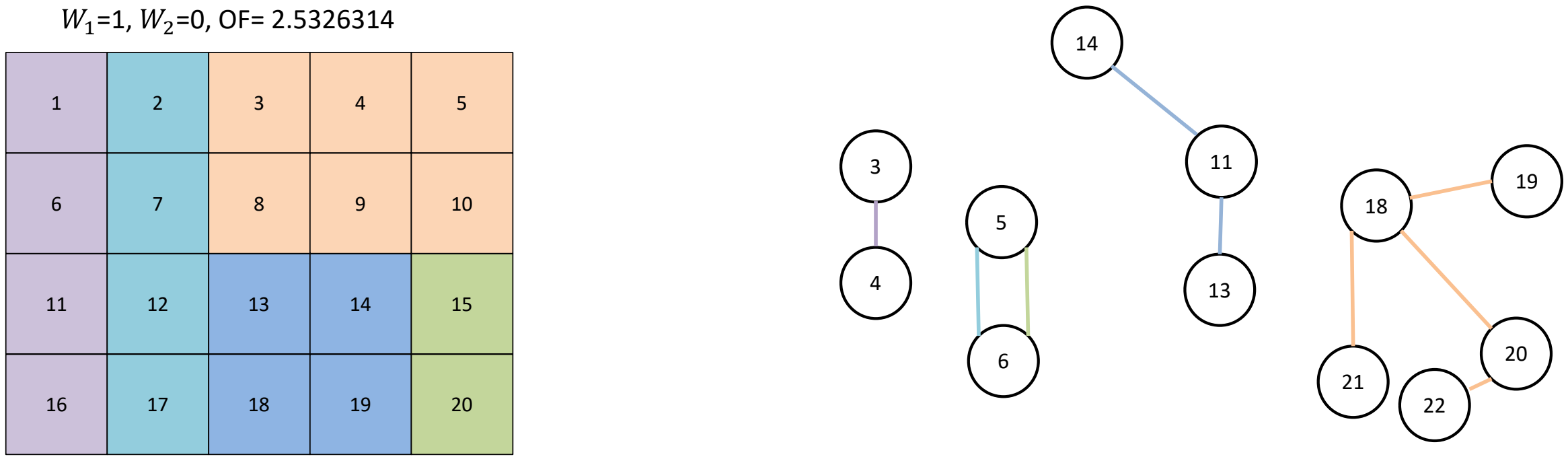

**(a)** Optimal region assignation. **(b)** Industrial interaction per region.

**Figure 6:** This figure summarizes the output of the optimization for this example. The five shaded areas in the first part correspond to the optimal regions, while the color of the links in the second part indicates the regions in which such interactions take place.

In the second example we illustrate the effect of the second term. We assume a simpler scenario where the country has nine areas with a corresponding decaying factor given by $\alpha = 2.4094$ and faces a PS as the depicted by Figure 7b. In this case we assume that the country produces four goods as labeled in Figure 7a so that in the purple areas goods 3 and 4 are produced with the same intensity, goods 5 and 6 are produced with the same intensity in the green areas and good 11 is produced in the white area. We assume that the links in the PS, particularly the one between goods 5 and 11 and the one between goods 3 and 11 are equal. Under this scenario, as we saw in the previous example, the purple and green regions should be preserved by the first term of the objective function. Our attention is focused now in how the white area in the middle is assigned to one of the two regions. Notice that the interaction between goods 5 and 11 pursues the arising of good 7 that is more complex than good 2. Thus, we expect that a solution of the $p$-IE model with $p = 2$ should assign good 11 to the green region, which effectively happens as we can see in the Figure 7c.

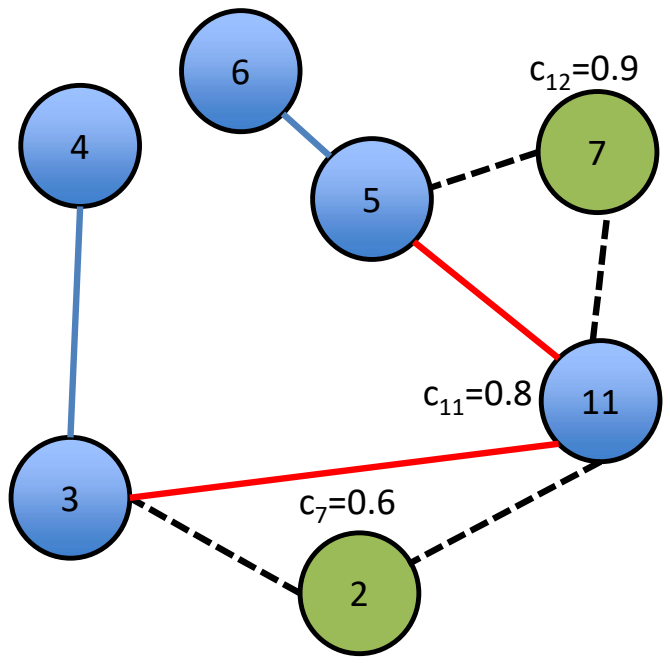

**(a)** PS example 2.

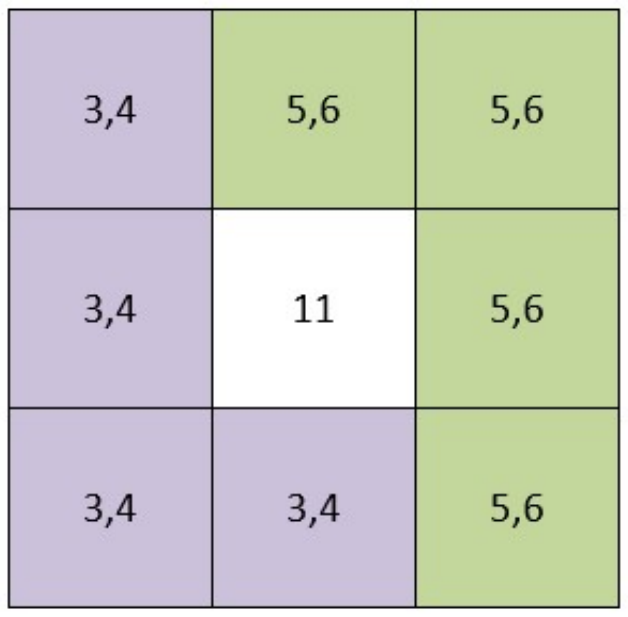

**(b)** Spatial distribution of the industries in example 2.

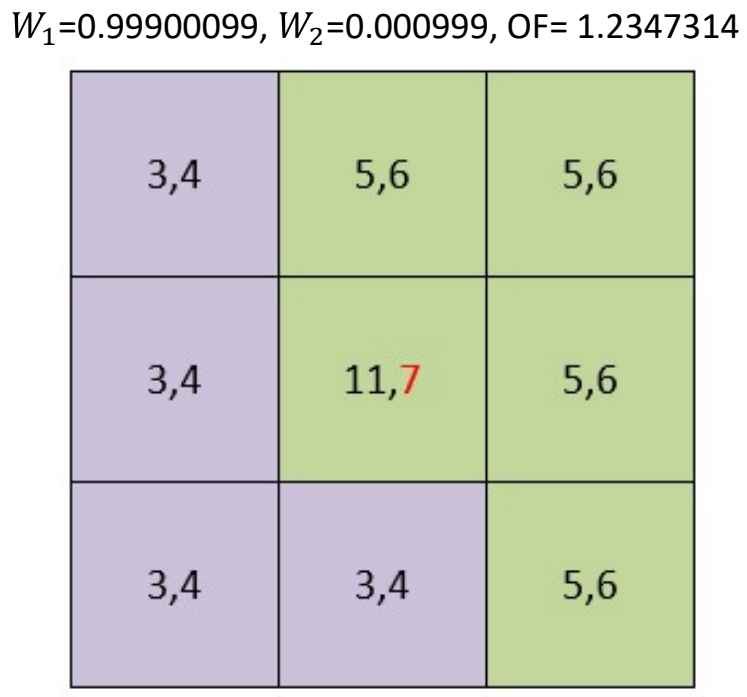

**(c)** Optimal region assignation.

**Figure 7:** This figure summarizes example 2. In Figure 7a we have the PS for this economy with 5 available nodes and two targets with their respective complexity. Figure 7b shows the spatial distribution of the available goods (3,4,5,6,11) which we assume are equally produced and divided into two spatial clusters: the one in which 3 and 4 interact (purple) and the one in which 5 and 6 interact (green). Figure 7c presents the optimal solution for example 2, depicting how the area containing 11 is assigned to the cluster generated by 5 and 6 enhancing the arising of 7 (in red).

# 4 The heuristic

In this section, we address the computational complexity of the $p$-IE problem by designing a heuristic solution. The use of heuristic methods has been a recurrent alternative within the family of the $p$-regions models (Rosenthal and Strange, 2004; De Groot et al., 2016; Laura et al., 2015; She et al., 2017). Regardless whether it is a $p$-regions or a Max-$p$-regions problem the structure of a heuristic for region design consists of two main blocks, (1) generate a set of initial feasible solutions and (2) take the best initial feasible solution and improve it by using a local search process based on tabu search, simulated annealing, greedy algorithm, among others (Rosenthal and Strange, 2004;Duque et al., 2007). Computational experiments made by Rosenthal and Strange (2004) show that the tabu search performs significantly better within the family of the $p$-regions problems. The use of tabu search on an initial feasible solution is a practice that dates back to the nineties with the seminal contribution by Opeshaw and Rao (Openshaw and Rao, 1995).

Algorithm 4.1 presents the heuristic for the $p$-IE problem. The construction phase is inspired on the "growing regions" strategy first proposed by Vickrey (1961), which generates an initial feasible solution in two steps: (1) select a set of p areas, which are called "seed" areas, and (2) starts assigning neighboring unassigned areas until all areas are assigned to a region (i.e., a region "grows" around each initial seed). There exist many variants of both steps; in our case, we select the initial seeds using the $k$-means++ algorithm proposed by Arthur and Vassilvitskii (2006), which seeks a good, spread out, location of the initial seeds. The $k$-means++ has a probabilistic component that allows exploring different sets of initial seeds. Each seed area becomes a growing region around which the rest of the areas are assigned. For the second step, the algorithm identifies the bordering areas (unassigned areas that share a border with a growing region) and selects the one with the minimum distance to its neighboring growing region. The distance between a bordering area $i$ and a growing region GR is calculated as the sum of squared differences of the vector of exports $r_{li}$ and the average vector of exports of the areas already assigned to the growing region GR (see equation (16)).

$$d_{iGR} = \sum_{l} (r_{li} - \text{avg}(r_{lGR}))^2. \tag{16}$$

Each time an area is assigned to a region it is necessary to update the set of neighboring unassigned areas and the process repeats until all areas are assigned to a region. Since the construction of an initial feasible solution is a fast process, we generate multiple initial feasible solutions (*maxitr*) and choose the one that maximizes the objective function $Z(\cdot)$. The best solution, $P_p^{\text{best}}$, is then passed to the next block, the local search.

The second block, the local search, uses a tabu search algorithm (Glover, 1977; Glover, 1989 and Glover, 1990), which has shown good performance in spatial aggregation models (Rosenthal and Strange, 2004; Openshaw and Rao, 1995). Given a feasible solution, the neighboring solutions, $N^s$, are obtained by moving bordering areas (i.e., areas that share a border with a neighboring region) to neighboring regions, one at a time, while preserving feasibility. The tabu search explores these neighboring solutions seeking improvements in the objective function $Z(\cdot)$. One key aspect of the tabu search is that it allows for a possible worsening of the objective function as a strategy for escaping from local optima with the hope that it will lead to an even better solution. The search stops after a predefined number of non-improving moves (*convTabu*). Although there exist multiple variants of this algorithm we decided to use the simplest version because the local search is computationally intensive and previous literature in the area of spatial clustering did not find clear advantages of using more sophisticated version of the tabu search algorithm (Rosenthal and Strange, 2004; Openshaw and Rao, 1995).

---

**Algorithm 4.1:** THE-P-INNOVATION-ECOSYSTEMS (
$A$ : Set of areas,
$p$ : Number of regions,
$maxitr$ : Number of initial feasible solutions to generate,
$r_{li}$ : Production of good $l$ in area $i$,
$lengthTabu$ : Length of the tabu list,
$convTabu$ : Number of non-improving moves before stop.)

---

**Comment:** Aggregate $n$ areas into $p$ spatially contiguous regions such that $Z(S_p)$ is maximized.

$P_p^{best} = \emptyset$, best partition.
**Construction Phase: Grow regions from seed areas**
**for** $i = 1, 2, \cdots, maxitr$
- **do**
  - $seeds =$ **k-means++**$(A, p, r_{li}, W)$
  - $S_p =$ **GrowRegions**$(seeds, A, r_{li}, W)$
  - **if** $Z(S_p) > Z(P_p^{best})$
    - **then** $P_p^{best} = S_p$

**Local Search Phase: Tabu search**
$P_p^* = P_p^{best}; P_p^{current} = P_p^{best}; tabuList = \{\}; c = 1$
**while** $c \leq convTabu$
- **do**
  - $N^s$ = Set of feasible neighboring solutions of $P_p^{current}$
  - **if** $N^s = \emptyset$
    - **then** $c = convTabu$
    - **else**
      - **for** $P_p^{new}$ in $N^s$
      - **do**
        - **if** $P_p^{new} \notin tabuList$
        - **then**
          - **if** $Z(P_p^{new}) < Z(P_p^*)$
          - **then**
            - $P_p^* = P_p^{new}$
            - $P_p^{current} = P_p^{new}$
            - $c = 1$
            - $N^s = \{\}$
            - $tabuList.add(P_p^{new})$
          - **else**
            - $P_p^{current} = P_p^{new}$
            - $convTabu = convTabu + 1$
            - $N^s = \{\}$
        - **else**
          - **if** $Z(P_p^{new}) < Z(P_p^*)$
          - **then**
            - $P_p^* = P_p^{new}$
            - $P_p^{current} = P_p^{new}$
            - $c = 1$
            - $N^s = \{\}$
            - $tabuList.add(P_p^{new})$
          - **else**
            - $N^s = N^s - P_p^{new}$
            - $tabuList.pop()$

**return** $(P)_p^*$

---

To illustrate the way in which each block of the tabu search contributes to finding a good solution we present the result obtained for the example in Figure 6 which aggregates twenty areas into five regions with an optimal objective function $Z = 2.5327$. To test each component (construct initial feasible solutions and local search), we ran the tabu search twice with different configurations: the first test was based upon generating 500 initial feasible solutions with the hope of finding a very good initial solution before moving to the local search. In the second test only two initial feasible solutions were generated. This second test relies on local search to improve the initial solution as much as possible (hopefully to reach the optimum). In both cases, we stopped the tabu search after ten non-improving moves (*convTabu*=10). Figure 8 shows the results of both configurations. When we explore 500 initial feasible solutions (Figure 8a) the algorithm finds a solution that takes only four iterations of the local search to find the optimal solution (plus ten non-improving moves for convergence). When we generate only 2 initial feasible solutions (Figure 8b) the solution that enters the local search is considerably further from the global optima compared to the previous configuration. In this case, the local search process requires thirteen iterations to find the optimal solution (plus ten non-improving moves for convergence). Note also how the local search process escapes a local optimal between iterations 8 and 13 in Figure 8b.

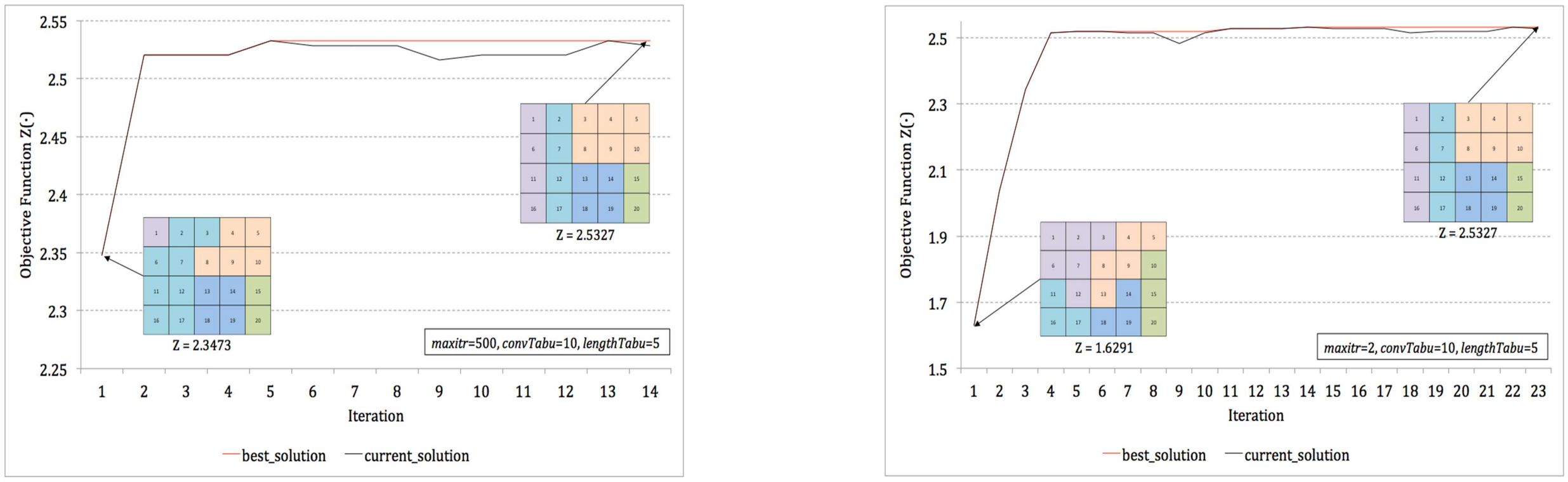

**(a)** 500 initial feasible solutions.

**(b)** 2 initial feasible solutions.

**Figure 8:** Performance of Tabu Search under two different configurations, in Figure 8a with a high number of initial feasible solutions, and in Figure 8a with a low number of initial feasible solutions.

# 5 Case of study

As an application of the $p$-IE model we address the problem of finding optimal industrial ecosystems for Colombia.

## 5.1 Data

*PS data*

We developed a version of the PS using trade data from the Centre dÉtudes Prospectives et d'Informatio Internationales (CEPII), which contains data for 128 countries, for the time period 1995 to 2010 and includes observations of 1,240 products classified under the nomenclature of the Harmonized System at the 4 digit level (HS4). We computed the weights of the network using equation (1). Following the lines of Hidalgo et al. (2007) we opt for a threshold of 0.55 to define the PS, although there is a range of possible thresholds that would work for our purposes. Also, since our dataset differs from the one used for the computations in Hidalgo et al. (2007), we provide in Figure 9 the respective analysis and diagnostics to justify why this election is still reasonable in our case. In our case, we select a PS with 708 nodes that includes the basket of exports of Colombia. Another important difference is that we have higher average degree centrality (6.823 vs 4 of the PS in Krugman (1991)). In our case, there are 100 nodes outside of the giant component. Nevertheless, all of them are in little components with negligible size. Thus, the connectivity and sparseness properties are basically preserved by the giant component and we can restrict our analysis to the latter one.

For a better visualization of our version of the PS and for the subsequent descriptive analysis of our results we identified 7 main clusters on the graph using the standard Louvain Modularity algorithm (Blondel et al., 2008). These clusters contain 92.92% of the nodes of the giant component. We can also identify the position of Colombia on the PS by detecting the set of nodes in which Colombia has RCA greater than 1. We summarize this information in Figure 10. Colombia reveals comparative advantage in 217 goods and 187 belongs to the giant component of the PS.

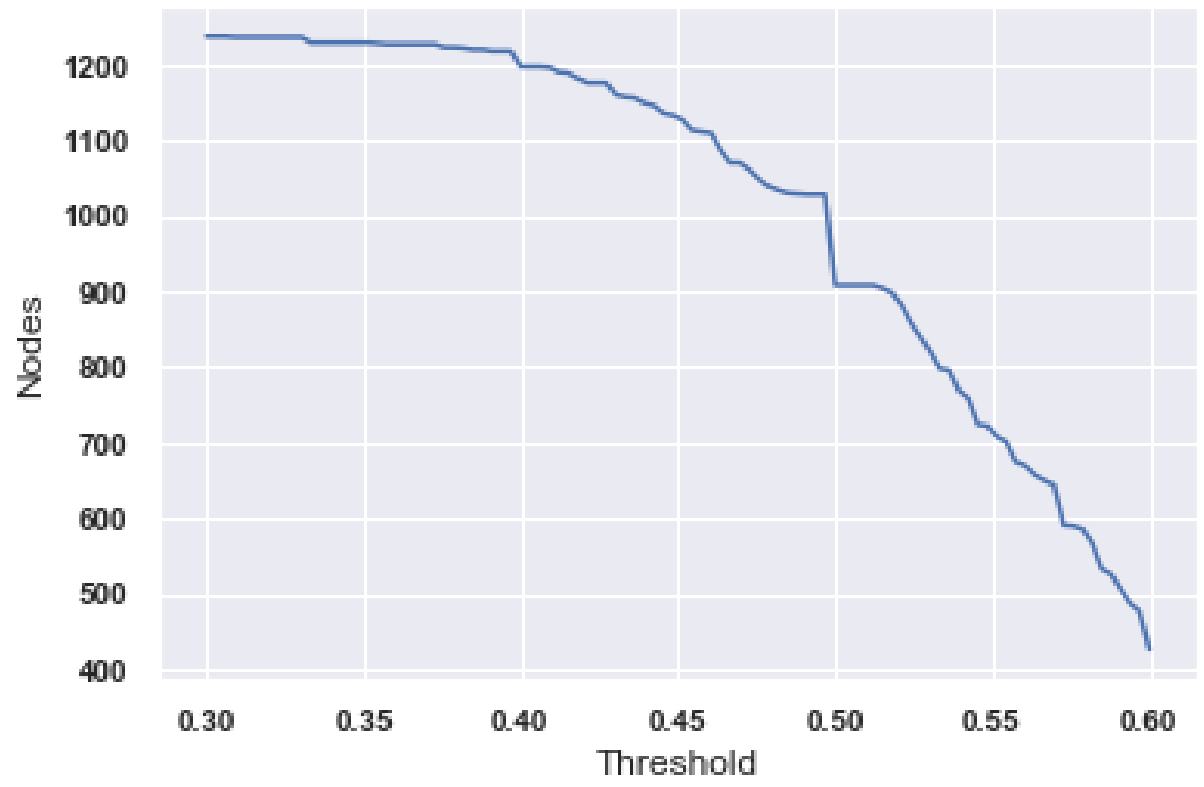

**(a)** Number of nodes in the PS vs value of the Threshold.

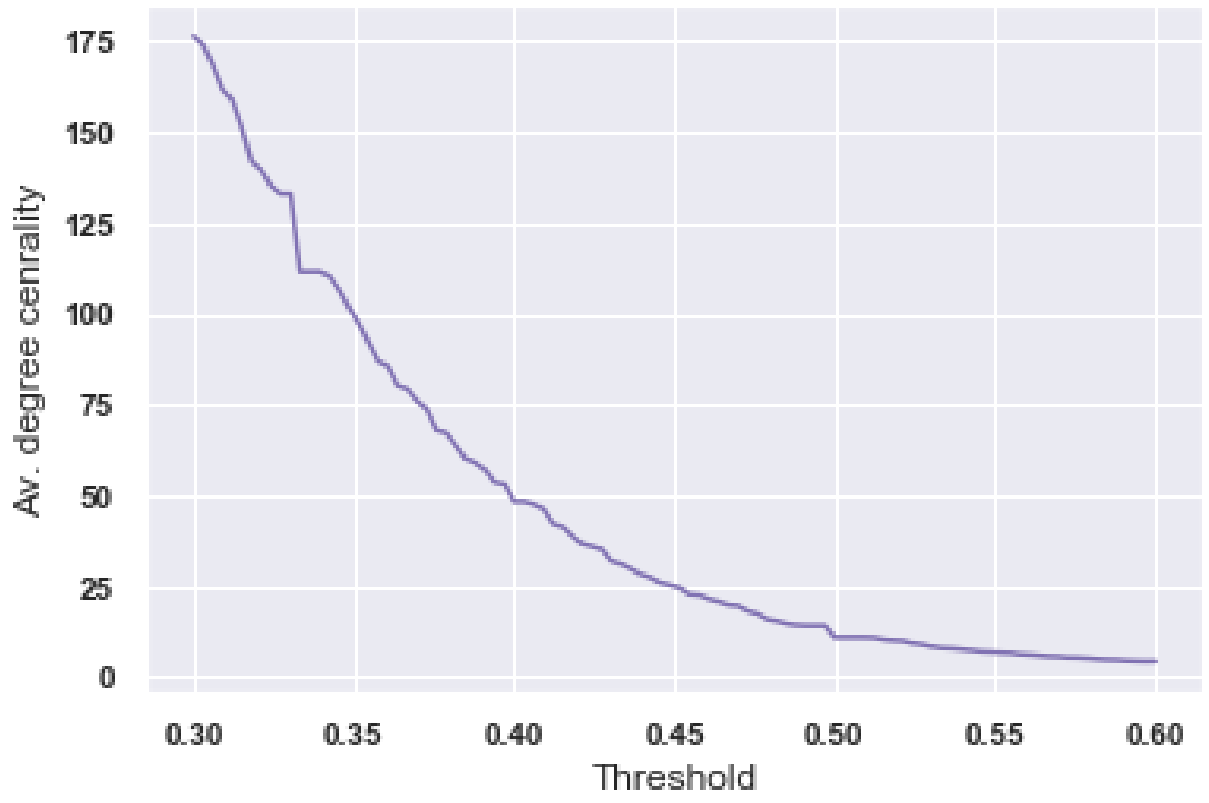

**(b)** Average degree of centrality in the PS vs value of the Threshold.

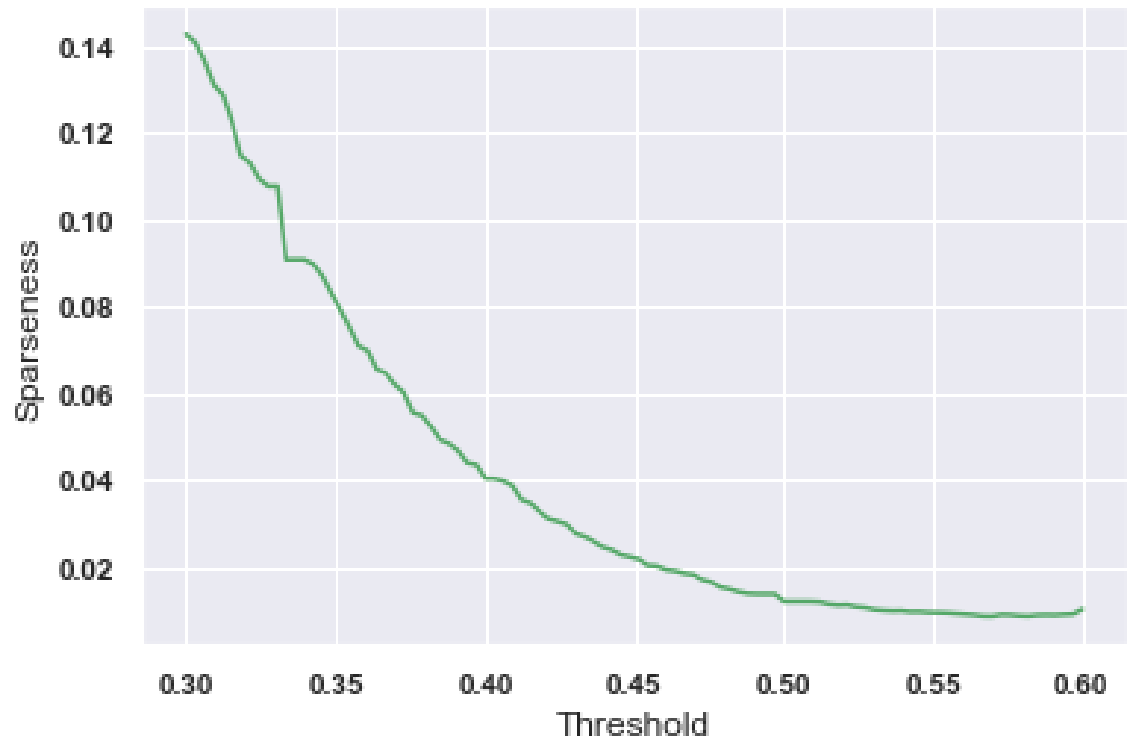

**(c)** Sparseness of the PS vs value of the Threshold.

| Metric (T=0.55) | Value |
|---|---|
| Number of nodes | 708 |
| Sparseness | 0.000027 |
| Av. degree centrality | 6.823 |
| Size giant component | 608 |
| Number of components | 39 |
| Av. size of small components | 2 |

**(d)** Description of the PS when a threshold of 0.55 is used.

**Figure 9:** We present how the network is becoming more informative as we increase the threshold. This is reflected in how the number of nodes in Figure 4a decays slowly, while the degree centrality in Figure 4b and the sparseness in Figure 4c decrease rapidly. With a threshold of 0.55 we have a significant number of nodes (708), most of them in the giant component and with small components of a negligible size (2 nodes in average) as it is showed in Figure 4d.

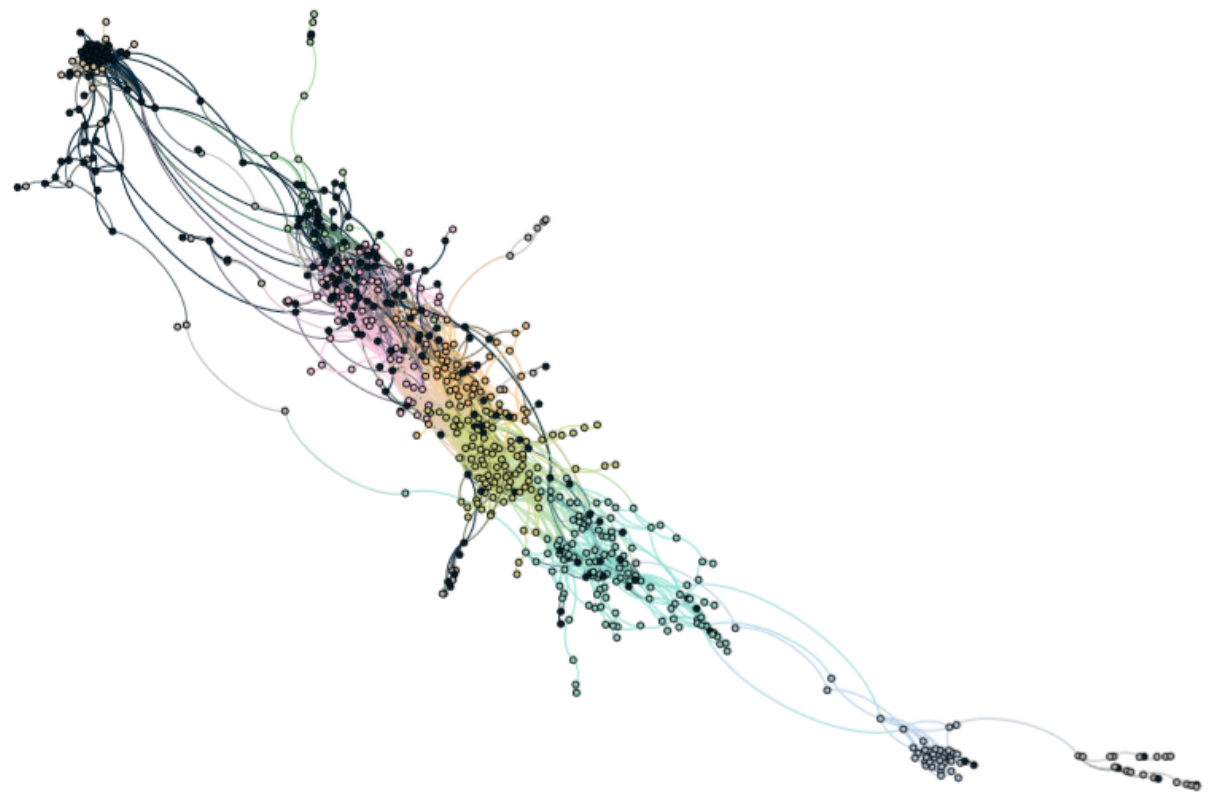

**(a)** Location of Colombia on the PS.

| Color code | Cluster | Share |
|---|---|---|
| | Clothing and fabrics | 16.04% |
| | Metals | 10.16% |
| | Chemicals | 32.62% |
| | Industrial metals and glasses | 5.88% |
| | Machinery & Electrical | 7.48% |
| | Electronics | 1.07% |
| | Agricultural products | 11.23% |
| | Other | 15.52% |

**(b)** Share of Colombian nodes in the PS.

**Figure 10:** In the first part we depict the communities detected with the Louvains method in the giant component of the PS and we mark in black, on top of them, the nodes in which Colombia reveal comparative advantage. The color code of the communities and the share of Colombia in each one of them is summarized in the second part. The order in which the communities are listed corresponds with their relative size compared with the giant component of the PS.

*Geography data*

The georeferenced data required to compute the interactions between industry areas is obtained from the open source information in Datlas tool of Bancoldex. This dataset contains the exports per industry in HS4 and per municipality for the years 2008-2017. We use this export data for the year 2014 to compute the distribution matrix (analogous to the one depicted in Figure 2b) and the subsequent parameters required for the $p$-IE model. Since several municipalities in Colombia have a very incipient or nonexistent industry, we found zero exports for 591 out of 1120 municipalities. We overcome this issue by merging municipalities with zero exports with neighboring municipalities with positive exports in at least one industry. We defined these new areas by implementing the Max-$p$ Regions algorithm (Duque et al., 2012), using exports as the decision variable. This algorithm merges municipalities into an optimal number of spatially continuous areas, following a queen adjacency criterion, so that the distance between municipalities is minimized. The distance function is defined as the Euclidean difference of the coordinates of the centroids of the municipalities plus the difference of the department ID of each municipality (the departments in Colombia are regional divisions similar to the states in USA). This distance is chosen such that the resulting regions are compact and prefer to remain within the same department (see Figure 11), preserving geographic cohesion. Finally note that the Max-$p$ Regions algorithm creates new areas such that only one of the member municipalities have positive exports.

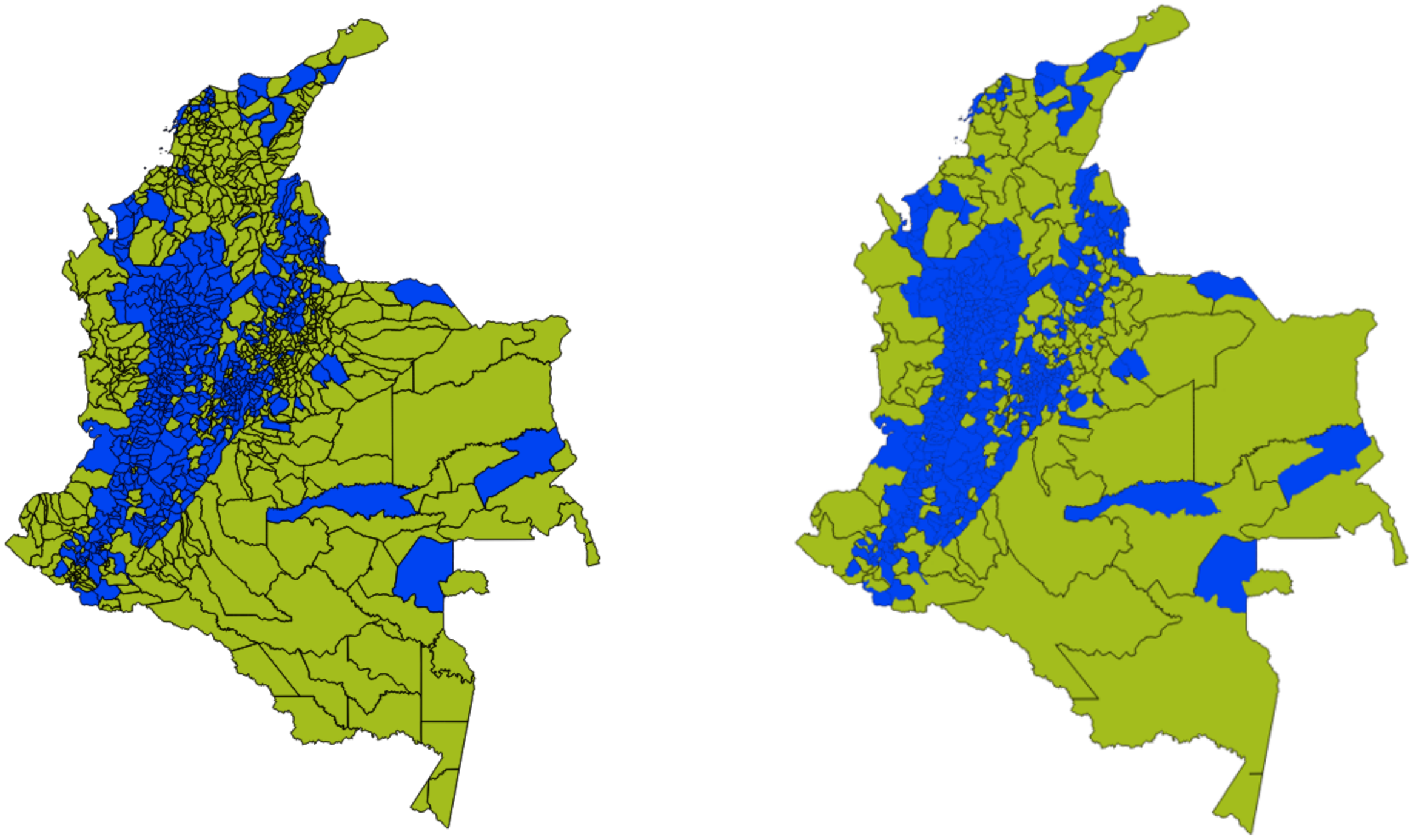

**(a)** Municipalities with positive exports.

**(b)** Areas obtained after applying the Max-p regions model.

**Figure 11:** The first part shows the map of Colombia divided in 1,120 municipalities, where 529 of them (marked in blue) have positive exports in some industry. Second part depicts the regions obtained after applying the Max-$p$ Regions algorithm to the first part.

The distance matrix $(d_{ij})$ is computed as the Euclidean distance between the centroids of the resulting 529 areas. We determined the decay parameter  similarly to the case of the examples in Section 3.4. In the same manner as that case we will take the parameter such that the distance decaying function $f$ vanishes at the radius of an average region of size $\frac{n}{p}$. However, since the geography in this case is less rigid, we assume that the optimal regions could resemble more a circle than a square. Hence, we will take $R_{av} = \sqrt{\frac{n}{\pi p}}$ as the radius of an average region and $M = 1,382.2$ km. Therefore, we have that $\alpha = \frac{\ln(2)}{\ln\left(\frac{M}{R_{av}}\right)}$.

*Results*

We solved the $p$-IE model using our heuristic for a range of values of $p$ between 10 and 90 as depicted in Figure 12. Since the heuristic algorithm (at least in the first stage) behaves like a clustering algorithm we opted for a simple application of the Elbow method to select a first candidate for the optimal number of regions. From this analysis we determined 40 as our first candidate.

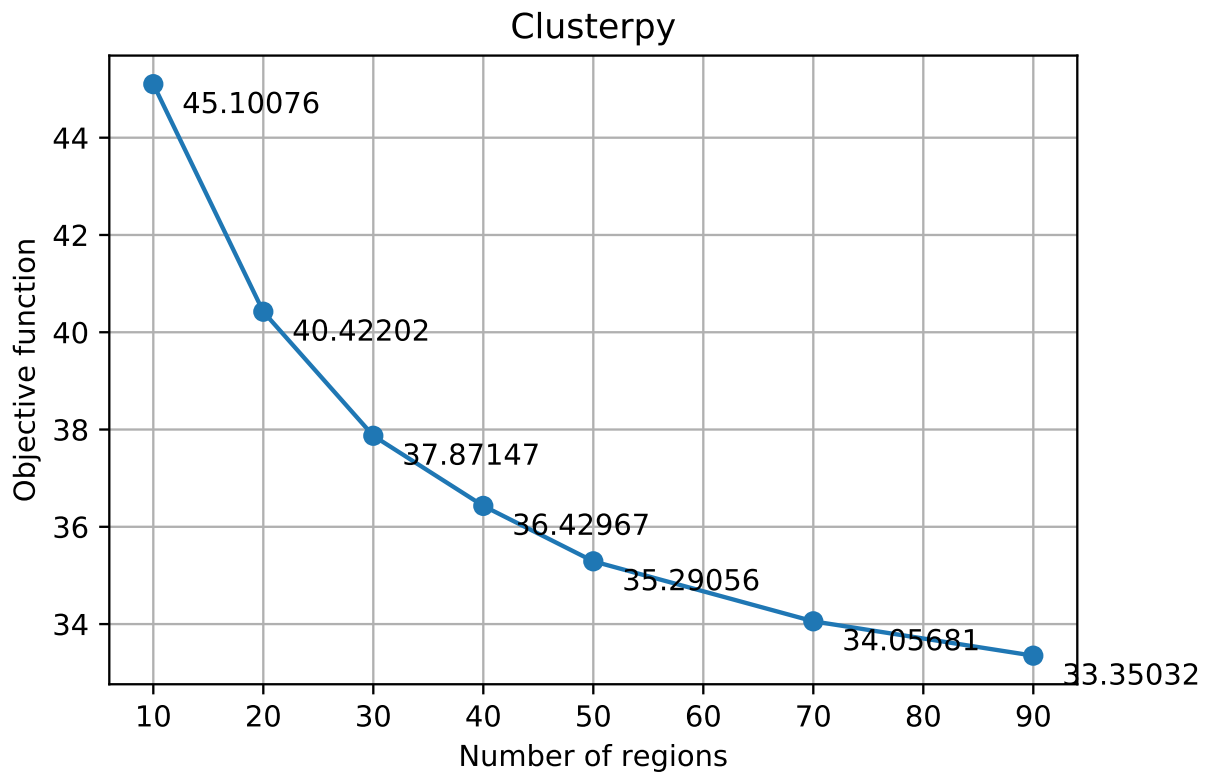

**(a)** OF vs Number of regions.

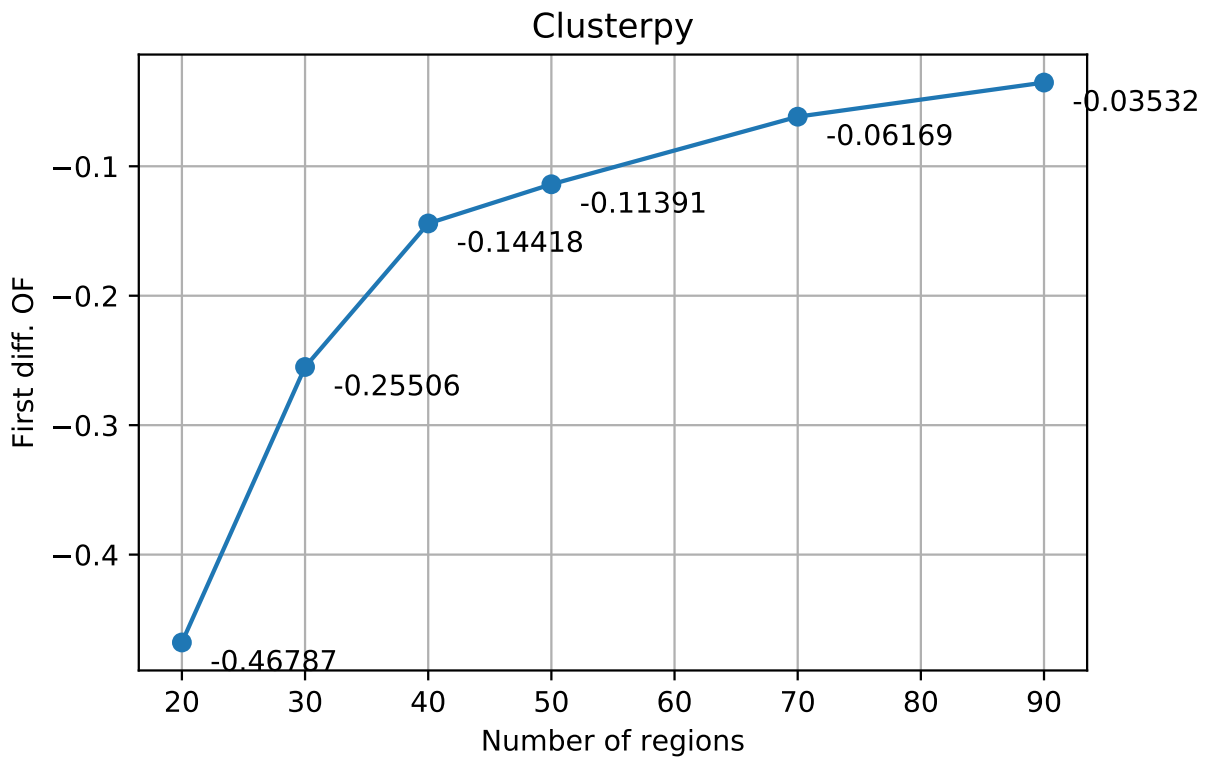

**(b)** First difference OF vs Number of regions.

**Figure 12:** In the first part we show how the optimal value of the objective function behaves in the number of regions and depicts how the slope starts to flatten significantly for big values of $p$. In part two we depict the behavior of the objective function, detecting that the biggest variation in the slope occurs at $p$=40.

However, the solution corresponding to $p$=40 has 15 singletons (regions constituted by one area). Different implementations of our model for smaller scenarios suggest that when the number of regions is high, the model has the tendency to isolate areas with less valuable interactions turning them into singletons so that the most valuable interactions are preserved within the bigger regions. This and several discussions with policy makers willing to implement this model in the Colombian case made us opt for a regional division without singletons so that we proposed to reduce the number of regions such that each singleton is incorporated into a bigger region. For this reason, we propose the model for Colombia with $p$=25. For this value of $p$ we have a decaying power $\alpha$ = 0.20041 and a configuration as depicted by Figure 13. We classify these regions according to their share in the total amount of exports of Colombia during 2014. We chose this classification to be consistent with the metrics of localization and of the PS that were based upon the export data. Table 1 provides an initial description of the geographic characteristics of the regions and the corresponding export share for each one of them.

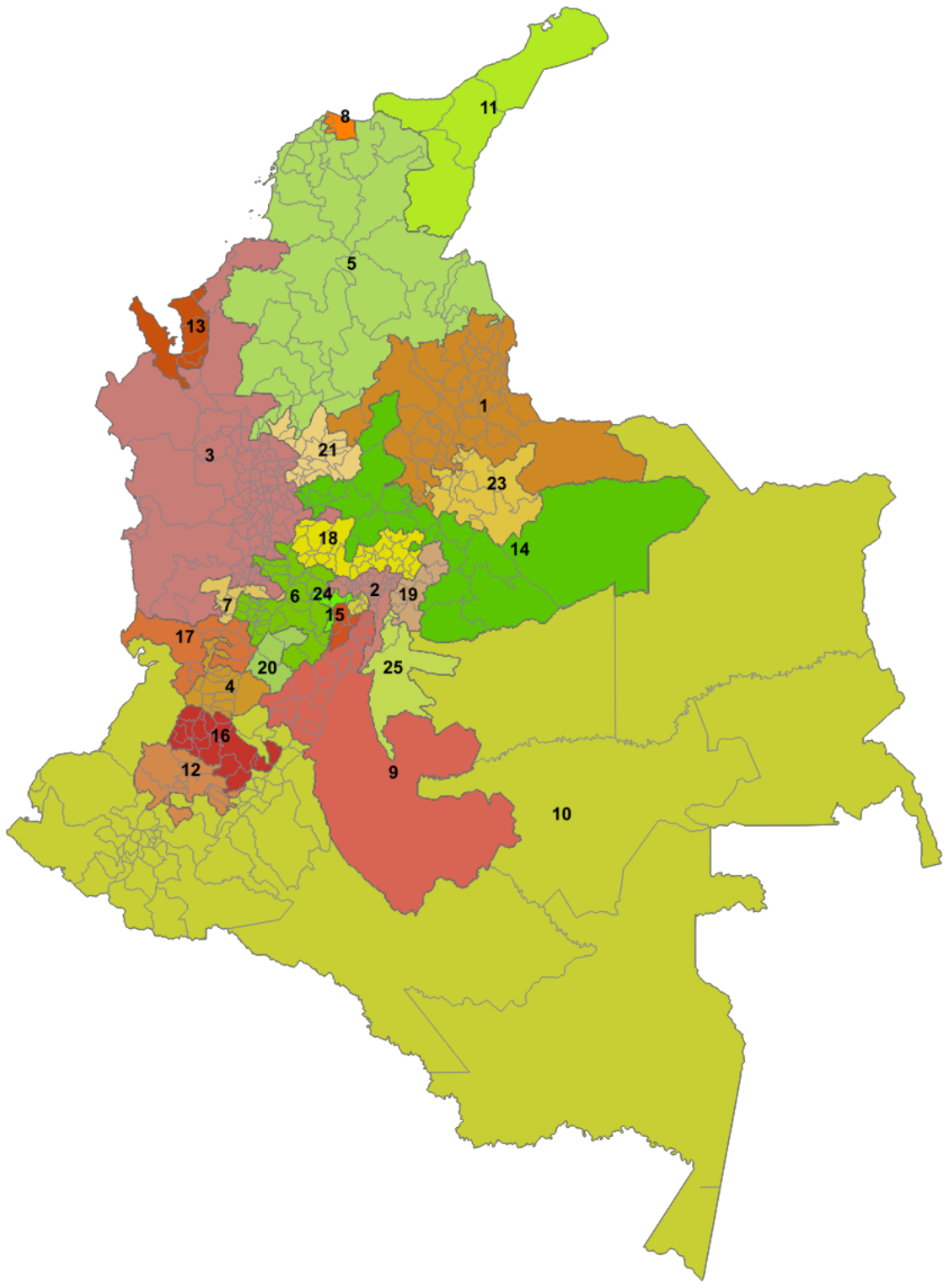

**Figure 13:** This figure depicts the solution for p=25 classifying the resultant regions with colors and labels. We label them in decreasing order with respect to their share in the total exports of Colombia in 2014 (see Table 1).

In Table 2 we summarize some relevant properties of the regions from the perspective of the Hausmann and Hidalgo's theory. The first two metrics correspond to a measure of the size of the subgraph spanned by each region within the PS. The column of industries shows how many industries of the giant component of the PS are produced in the region. The second column corresponds to the sparseness of each subgraph (see Barabási et al., 2016), i.e., for a region with $N$ industries and $L$ links among these industries its sparseness is given by $\frac{2L}{N(N-1)}$.

This also provides a sort of measure of the extent of interaction among key industries inside the region. The third metric indicates the relevance of the goods produced within the region as stepping stones in the PS. We compute the fraction of shortest paths between any pair of nodes passing through the nodes of the region, i.e., the group betweeness centrality for these nodes (see Brandes, 2001). This measure gauges the extent in which goods produced in the region are determinant in the diffusion processes on the PS. The fourth column is a weighted average of the normalized complexities of the goods produced within the region. More explicitly, for a region $R$ producing the set of goods $P$ (a subset of the PS) we consider the complexity measure

$$C_P = \sum_{i \in R} \sum_{l \in P} r_{il} c_l \tag{17}$$

| **Region ID** | **Export share** | **Number of areas** | **Area ($Km^2$)** |
|---|---|---|---|
| 0 | 19.123% | 48 | 54,760 |
| 1 | 19.049% | 28 | 4,958 |
| 2 | 18.565% | 102 | 85,578 |
| 3 | 10.650% | 16 | 7,095 |
| 4 | 6.619% | 40 | 102,746 |
| 5 | 6.173% | 36 | 13,755 |
| 6 | 3.895% | 8 | 2,774 |
| 7 | 3.894% | 3 | 1,298 |
| 8 | 2.309% | 21 | 70,797 |
| 9 | 2.136% | 58 | 369,591 |
| 10 | 1.787% | 4 | 33,649 |
| 11 | 1.679% | 13 | 9,673 |
| 12 | 0.896% | 5 | 7,349 |
| 13 | 0.828% | 23 | 74,130 |
| 14 | 0.714% | 4 | 1,627 |
| 15 | 0.430% | 18 | 9,394 |
| 16 | 0.417% | 11 | 10,406 |
| 17 | 0.306% | 29 | 10,762 |
| 18 | 0.208% | 10 | 5,051 |
| 19 | 0.207% | 3 | 4,199 |
| 20 | 0.066% | 20 | 8,430 |
| 21 | 0.026% | 6 | 687 |
| 22 | 0.013% | 14 | 14,621 |
| 23 | 0.008% | 5 | 1,166 |
| 24 | 0.002% | 4 | 11,152 |

**Table 1:** Summary of some initial characteristics of the resultant regions.

| **Region ID** | **Number of industries** | **Sparseness** | **Betweenness centrality** | **Complexity** |
|---|---|---|---|---|
| 1 | 73 | 7.116% | 0.40138 | 1.322 |
| 2 | 88 | 6.792% | 0.44730 | 15.509 |
| 3 | 89 | 6.639% | 0.44676 | 13.757 |
| 4 | 80 | 7.184% | 0.44494 | 9.441 |
| 5 | 78 | 7.426% | 0.43160 | 2.699 |
| 6 | 57 | 8.521% | 0.28186 | 0.580 |
| 7 | 57 | 10.338% | 0.24048 | 0.386 |
| 8 | 78 | 6.993% | 0.43553 | 3.465 |
| 9 | 43 | 6.755% | 0.29255 | 0.161 |
| 10 | 38 | 10.242% | 0.33938 | 0.189 |
| 11 | 25 | 8.333% | 0.24083 | 0.073 |
| 12 | 21 | 5.714% | 0.09762 | 0.170 |
| 13 | 41 | 13.537% | 0.19502 | 0.046 |
| 14 | 21 | 16.190% | 0.14960 | 0.020 |
| 15 | 74 | 6.479% | 0.37626 | 0.974 |
| 16 | 30 | 7.586% | 0.21120 | 0.151 |
| 17 | 40 | 7.692% | 0.24234 | 0.239 |
| 18 | 47 | 7.031% | 0.31164 | 0.076 |
| 19 | 20 | 8.421% | 0.26320 | 0.095 |
| 20 | 1 | 100.000% | 1.00000 | 0.002 |
| 21 | 55 | 8.350% | 0.34788 | 0.186 |
| 22 | 11 | 5.455% | 0.09763 | 0.008 |
| 23 | 5 | 10.000% | 0.06123 | 0.001 |
| 24 | 31 | 7.957% | 0.29851 | 0.068 |
| 25 | 5 | 0.000% | 0.04698 | 0.006 |

**Table 2:** Summary of some attributes computed for each region for the solution corresponding to $p$=25.

We want to point out that some strange features of this solution are a direct consequence of the Colombian export structure. For example, the large region 10 is formed by 58 areas where 10 of of the municipalities make up 70% of its size in PS; this reflects the absence of exports (or significant exports) in this part of the country and precludes us to define finer divisions. Another example is given by region 20. This region only exports one good in which Colombia has a comparative advantage (coffee beans), though this good has small betweness centrality and complexity values. These two radical cases show the main reasons that could force the model to deviate from expected solutions: gathering areas into macro-regions or defining regions specialized in goods with low complexity.

*Analysis of regions 1 and 3*

We conclude our analysis of the output with a comparison between region 1 and region 3. We choose these two for two reasons (1) both have a relatively similar size in $Km^2$; (2) the exports structure of both regions are distinct enough so that we can exemplify with them two types of regions in Colombia: regions following a diversification strategy (region 3) and regions specialized in agriculture or extractive economic activities (region 1).

In Figure 14 and Figure 15 we provide a thorough description of region 1 and region 3 that may be useful for policy making if we regard them as innovation ecosystems. We provide a detailed account of the exports of each region in the form of tree diagrams in part (c) of each figure. These diagrams are classified according to the communities identified and described in Figure 10b. It is worth noting that these communities do not follow any standard industrial classification, but are completely determined with clustering methods (Louvain algorithm). Thus, it is possible to have industries which do not seem to belong to the corresponding productive sector. The size of the boxes within each category are weighted by the betweness centrality (BC) of the industry as a node in the PS. As discussed before, the BC tells us how central a node is in the diffusion of a country on the PS. Actually, we may think of them as stepping-stones in the process of reaching more valuable or complex nodes. Thus, since region 1 and region 3 are intended to foster innovation processes over the short and long run, policy makers should focus not only on the complex goods, but in those which are essential to move towards them.

Region 1 is specialized in the refinement of petroleum and contributes 18% of the 44% of the share of the petroleum in Colombian exports. Even more dramatically, the origin of this export is concentrated in one city into region 1 (Barrancabermeja) which has one of the most important refineries of the country. On top of this, after taking the subset of goods that belong to the PS with the prescribed threshold (which excludes the petroleum) we see that region 1 is still specialized in the production of commodities such as melons and tomatoes (see Figure 14c). In contrast, region 3 has relatively strong industries contributing significantly in the production of complex goods (reaction & catalytic products, acrylic polymers, other rubber products, etc.) combined with the production of commodities (in the rural parts of the region) such as fermented milk, cheese and so forth (see Figure 15c).

The graphic information provided in parts (b) and (c) of Figure 14 and Figure 15 enables us to give a more detailed answer to the issue of determining the current situation of the region. In fact, from Table 2 we already know the average complexity, the number of industries and how central the goods of each region on average are. From Figure 14 and Figure 15 we can explain in what sense region 3 is more complex and more central (on average) than region 1. Since the boxes in the tree diagrams in Figure 14c and Figure 15c also contain information about the complexity of goods normalized between 0 and 1 (Comp) and the share of the country in total exports of each good of the country during that year (Sh), we can easily observe that products in region 3 account for both a large share of exports and a high complexity (in contrast with region 1). On top of this, the network of available and target goods rendered in part (b) of Figure 14 and Figure 15 allow us to see how likely existing industries will thrive. For example, we point out that in region 3 the clusters of Clothing and fabrics and Chemicals are bigger and better connected than in region 1. This is of particular interest in terms of complexity since the cluster of Chemicals in each region contains (in average) more complex goods than the other clusters.

Finally, we present prospective strategies of innovation for each region. Our implementation of the $p$-IE model finds 25 target nodes for Colombia. All 25 targets are described in Table 3 together with their respective complexity values and the probability of each of the two regions explored to introduce these new goods in their economy. These probabilities are computed as a weighted sum of the following type

$$\text{Prob Reg}_j(v) = \frac{\sum_{i \in R_j} \sum_{l \in P} r_{il} y_{lv}}{\sum_{l \in P} y_{lv}}. \tag{18}$$

where v is the target good, $R_j$ is the set of areas forming the region $j$ (for $j$=1,2). This computation is a linearized and continuous version of the formula proposed by Alshamsi et al. (2018). Since our computations use as inputs shares of the regions with respect to the total production of the country, these probabilities must be understood in a relative way, i.e., how likely a region is to develop comparative advantage in a good with respect to another. We notice that region 3 has a positive probability to branch into all 25 products, whereas region 1 only exhibits 21 products with positive probability. We can illustrate how to use this information with two simple examples. Combining parts (b) and (c) of Figure 14 and Figure 15 with Table 3 we notice that region 3 has a considerably higher probability of developing comparative advantage in the last 4 goods listed in Table 3 because all of them are highly related to goods that are intensively produced by region 3 and that belong to the cluster of Clothing and fabrics which is highly connected in this region. On the other hand, the most complex target good (fork-Lifts) has a considerably higher probability to be produced in region 3 compared with region 1. This difference underscores the fact that region 3 produces a significantly higher share of the products in the cluster of Metals related to forklifts in the PS as we can see in Figure 15b and Figure 15c, respectively.

| Name | Complexity | Prob Reg1 | Prob Reg3 |
|---|---|---|---|
| **Fork-Lifts** | 0.76808 | 0.095% | 14.164% |
| **Lubricating Products** | 0.74126 | 0.070% | 7.620% |
| **Transmissions** | 0.72552 | 0.101% | 4.427% |
| **Electrical Lighting & Signalling Equipment** | 0.70289 | 0.129% | 6.088% |
| **Vehicle Parts** | 0.69214 | 0.130% | 4.803% |
| **Large Flat-Rolled Iron** | 0.68408 | 2.518% | 4.728% |
| **Rock Wool** | 0.68167 | 1.619% | 3.608% |
| **Lifting Machinery** | 0.67676 | 2.320% | 0.310% |
| **Traffic Signals** | 0.67154 | 0.104% | 2.662% |
| **Iron Springs** | 0.67059 | 0.149% | 3.452% |
| **Locomotive Parts** | 0.66780 | 1.946% | 4.356% |
| **Rubber Pipes** | 0.65993 | 0.129% | 5.466% |
| **Engine Parts** | 0.64322 | 0.153% | 3.529% |
| **Metal Insulating Fittings** | 0.63526 | 0.245% | 9.380% |
| **Electric Motor Parts** | 0.62512 | 0.119% | 3.983% |
| **Glass Fibers** | 0.60791 | 0.001% | 0.017% |
| **Whey** | 0.58710 | 0.000% | 14.066% |
| **Electrical Insulators** | 0.58070 | 0.000% | 4.167% |
| **Letterstock** | 0.53960 | 0.000% | 7.697% |
| **Bovine Meat** | 0.50841 | 0.000% | 17.160% |
| **Tulles & Net Fabric** | 0.34832 | 0.719% | 8.273% |
| **Other Non-Knit Clothing Accessories** | 0.32376 | 0.054% | 15.447% |
| **Other Synthetic Fabrics** | 0.32241 | 0.376% | 27.458% |
| **Cotton Sewing Thread** | 0.31512 | 0.102% | 26.012% |
| **Textile Scraps** | 0.29533 | 0.602% | 60.167% |

**Table 3:** Target industries for the solution ranked by complexity. In here Complexity stands for the standardized complexity index of the goods (17), Prob Reg 1 and Prob Reg3 are the probabilities of region 1 and region 3 to produce such goods, respectively, computed using (18). All 25 have a positive probability to appear in region 3, nonetheless there are four industries that do not have any related activities in region 1 and correspond to the zeroes in the third column.

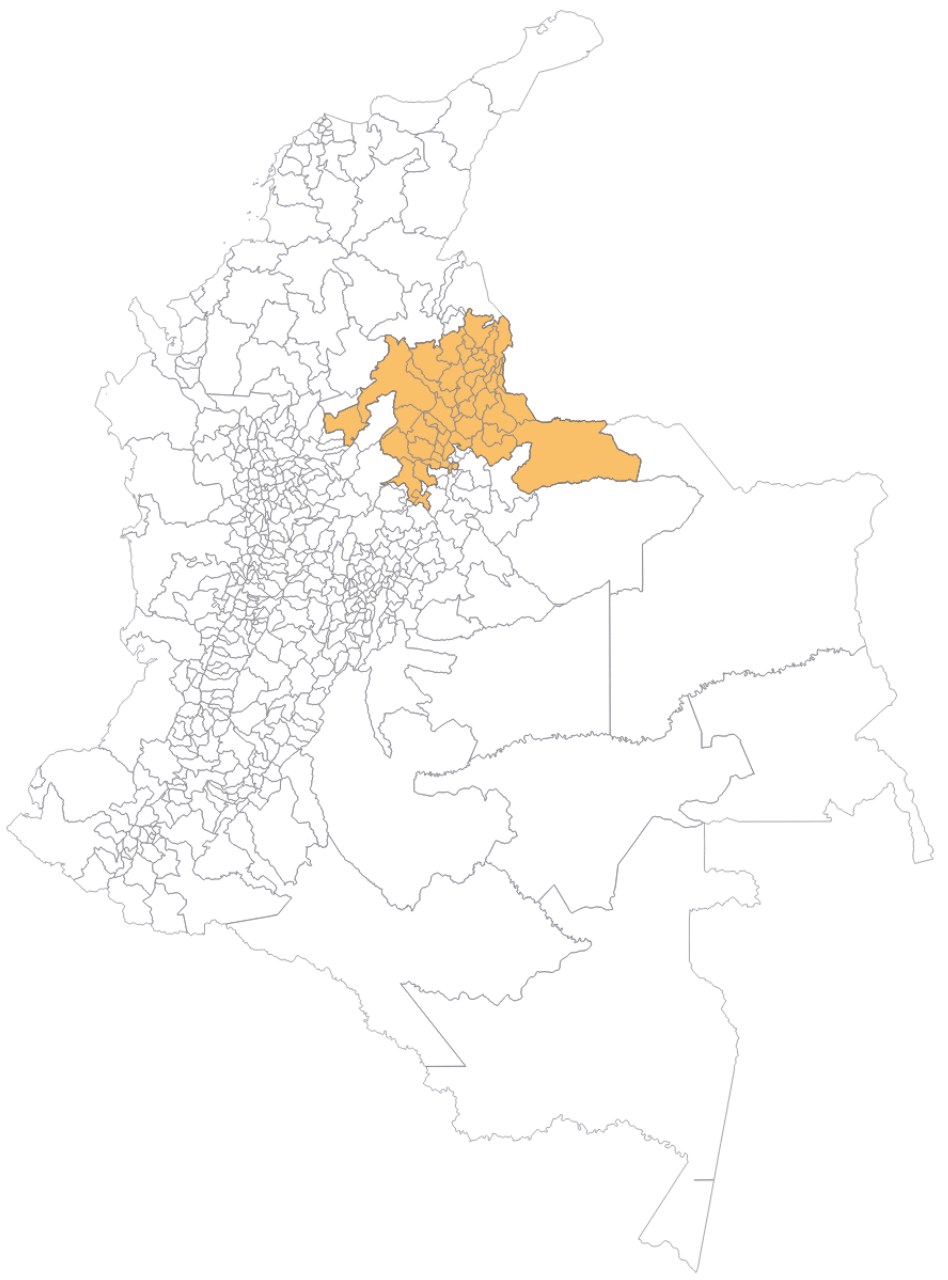

**(a)** Map of region 1.

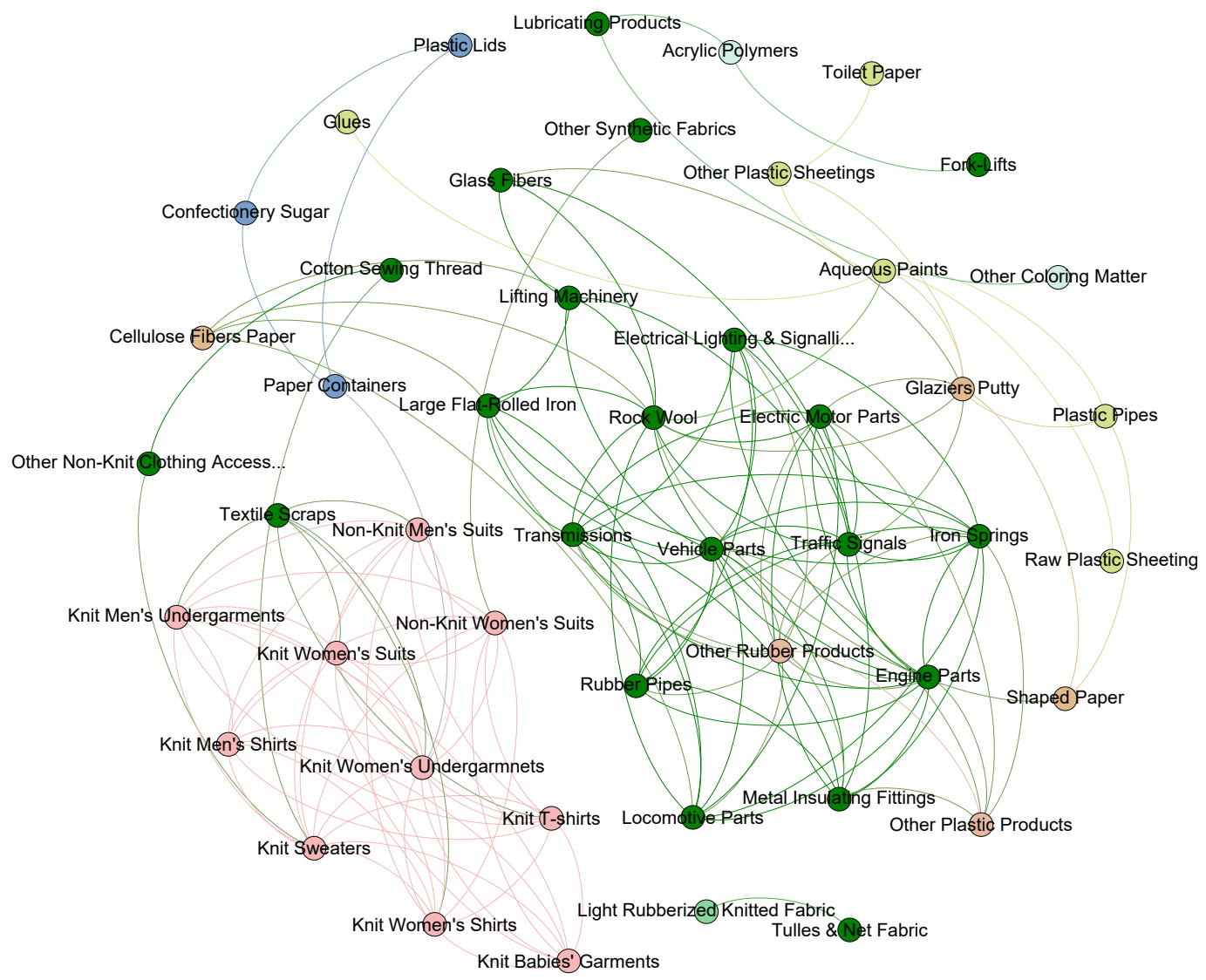

**(b)** Location of region 1 in the PS.

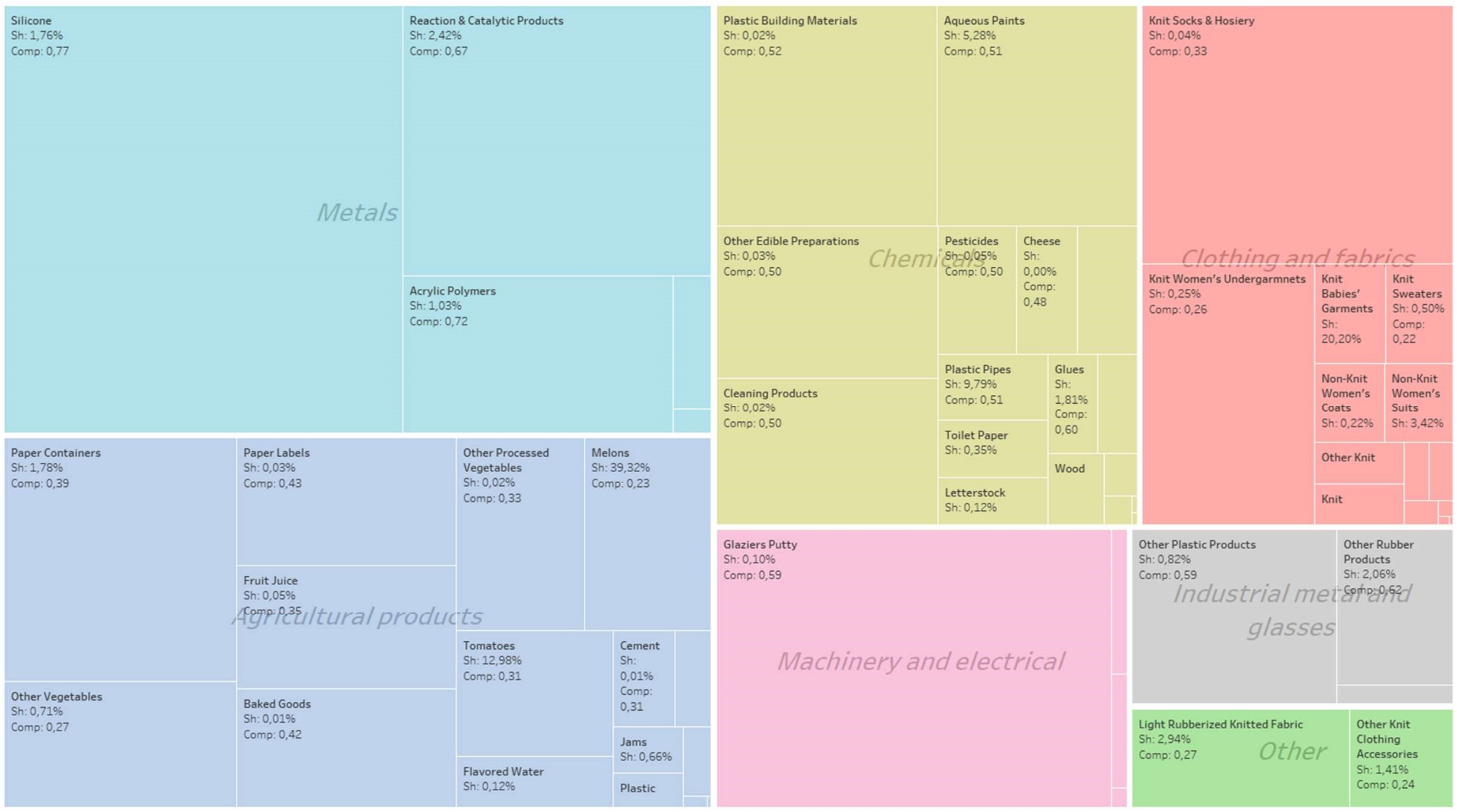

**(c)** Exports of region 1 per cluster.

**Figure 14:** Figure 14a depicts the geographic location and the areas that form region 1, Figure 14b depicts in green the target goods of region 1. All the other nodes correspond to available goods in the region colored according to the nomenclature established in Figure 14b. The box diagram depicted in Figure 14cto the betweenness centrality of the nodes in the PS and each box also contains the share of the region in that activity with respect to the national value (Sh) and the respective normalized complexity index (Comp).

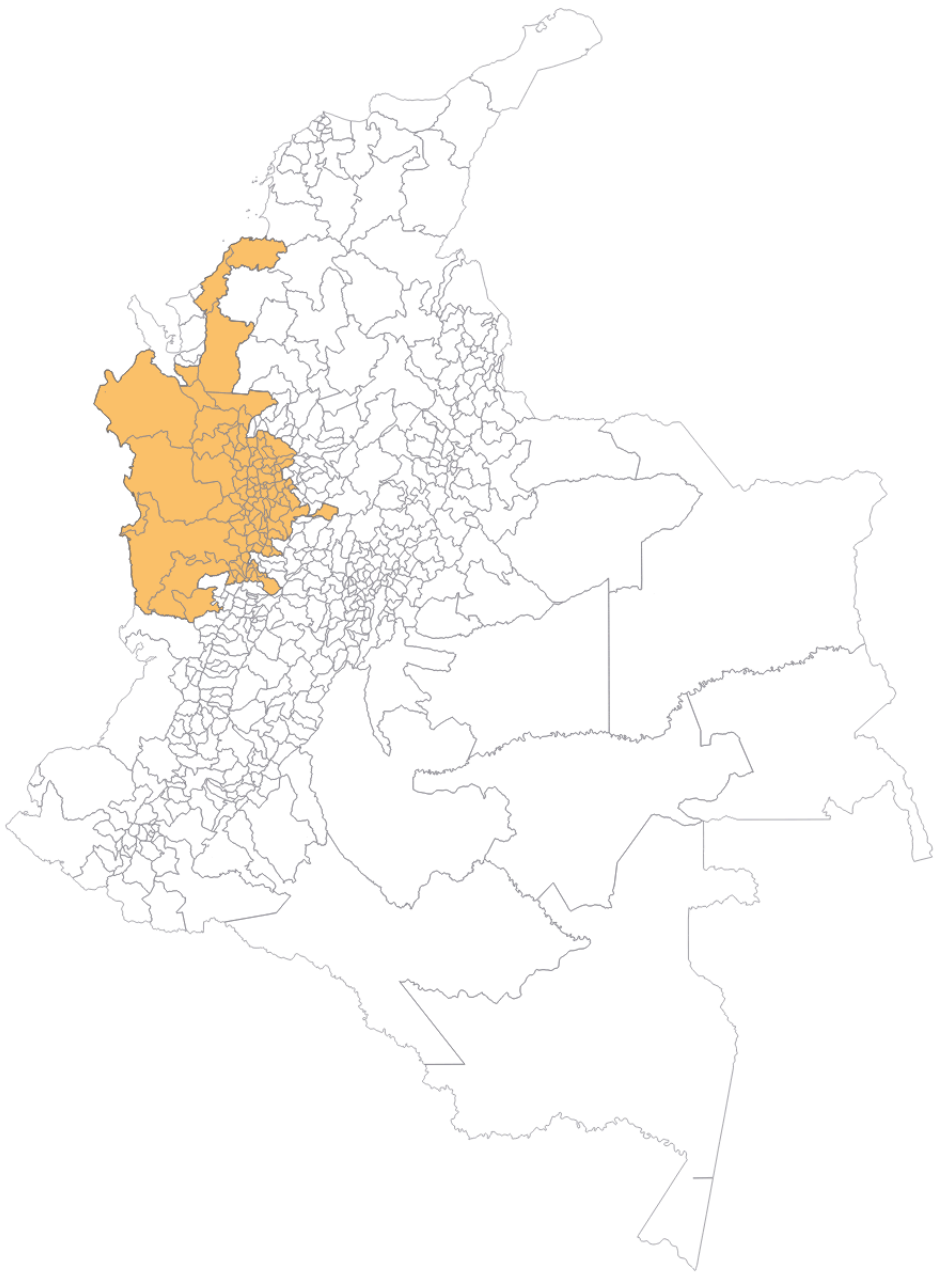

**(a)** Map of region 1.

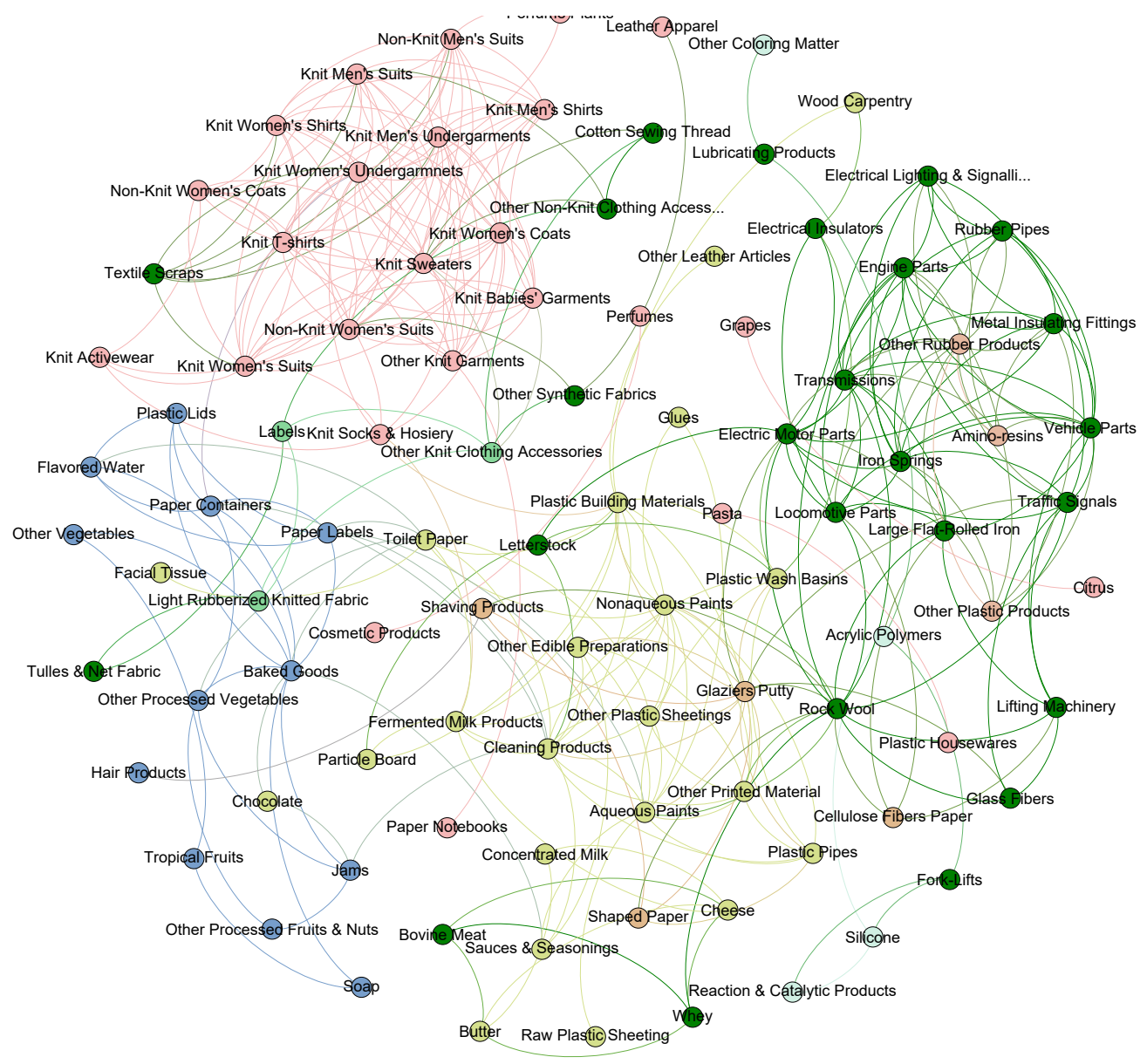

**(b)** Location of region 1 in the PS.

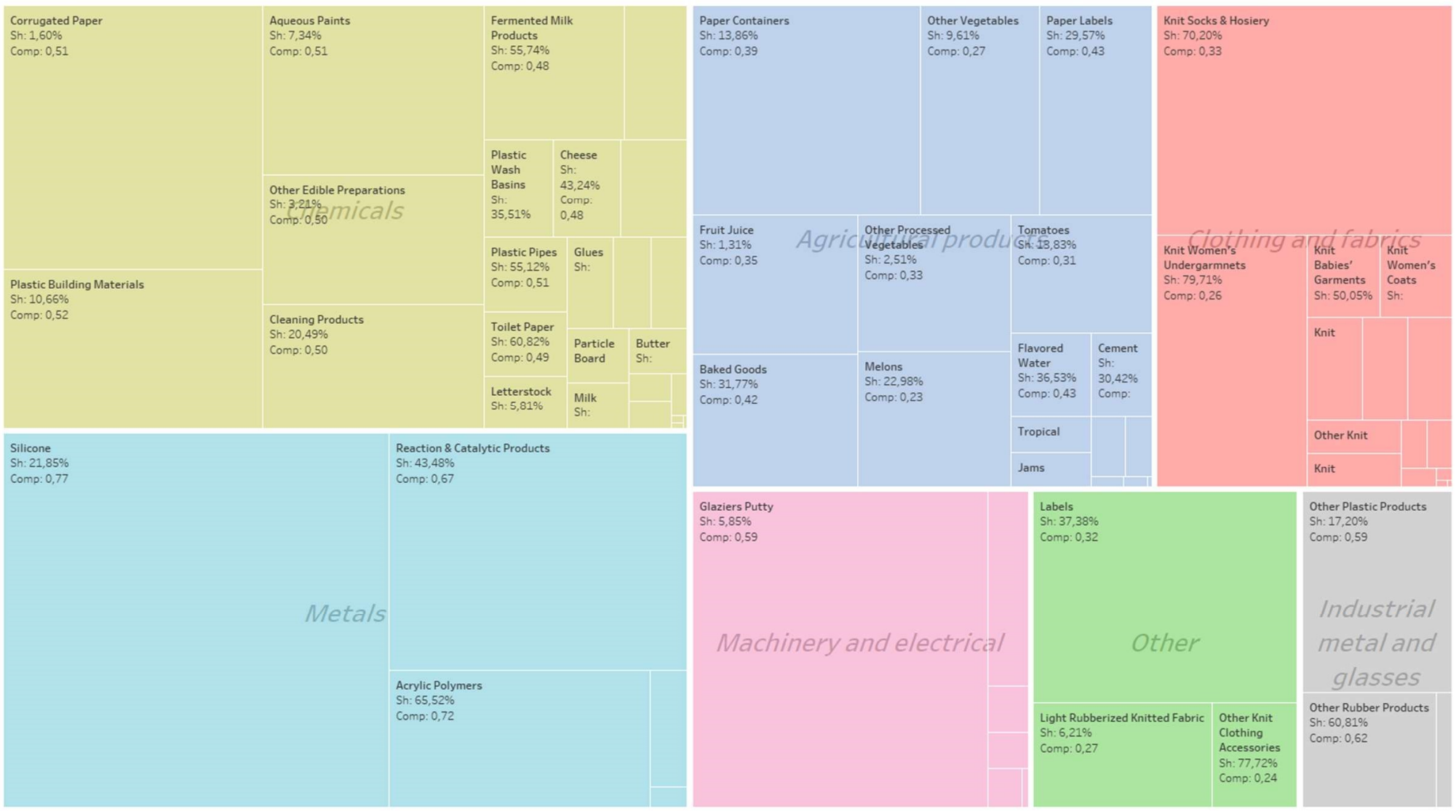

**(c)** Exports of region 1 per cluster.

**Figure 15:** This figure is analogous to Figure 14. We want to stress the fact that the PS of region 1 depicted by Figure 15b is a strict subset of the one depicted in Figure 14b. In particular, region 3 has a larger set of available and target nodes. Also, it is important to notice the change in the composition of the tree map in Figure 15c with respect to the one in Figure 15c.

# 6 Conclusions

In this paper we introduced a new member of the $p$-regions clustering methods called the $p$-Innovation Ecosystems (p-IE) model. This model adds a new layer of complexity into the set of $p$-regions models incorporating an underlying network structure that captures possible interactions between agents within the geographic areas. The main application (and motivation) of the model is the study of the evolution of the economic output based on recent developments of economic complexity and evolutionary economic theory. More precisely, this model consists in the aggregation of n areas into p spatially contiguous regions that (1) maximizes the number of relevant interactions among the industries within the same region; and (2) identify strategic relationships between industries within regions that maximize the probability of activating innovation processes that allow the region to jump to more complex relevant goods yet to appear in the economy. Formally, the $p$-IE model was formulated as MIP problem with a corresponding heuristic solution. From the economic point of view, we introduced this model as an application and extension of the theory of Hausmann and Hidalgo (done at the level of countries) to the study of the industrial evolution at regional levels.

From the perspective of policy development, we introduced several metrics as well as suggested how they might be applied in conjunction with the $p$-IE model. This approach will help policy makers diagnose and assess the actual performance of the economic regions and the potential performance of each of the regionalizations proposed by the model. We introduced these ideas in a case of study for the municipalities of Colombia and were able to identify 25 innovation ecosystems, which we ranked according to their contribution to the total exports of Colombia. We found a positive association between the complexity of these regions, their level of exports and the growth of their level of exports. Also, a further analysis applied to the first and third regions (according to our classification) showed how to obtain inputs that would help the policy makers design policies to exploit the current situation of the regions (amount of production) and to enhance the production of key industries in the innovation process (with high betweenness centrality) so that the target industries (with a high complexity) arise within the regions.

As it is formulated, the p-IE model can be implemented to other clustering problems as long as the situation can be described in terms of two inputs: (1) distribution of areas inside a region which contain agents or attributes of some sort, (2) a network structure that relates the agents or attributes identifying the most important attributes or agents and the key interactions among them. We suggest two alternative applications of this model. First, given an input-output matrix (of money transfers among industries) we could use the model to produce optimal regions so that the policy maker can identify the supply chains, preserve them within the resulting regions and design strategies to complete the chains inside the regions. Second, we can think in a scenario involving a network of amenities in a city (such as hospitals, schools, museums, gas stations, etc.) and the commuting flows or connectivity flows weighting the links among them. In this case, it would be possible to apply the model to divide the city into optimal zones in which a citizen could find well-connected and complementary sets of amenities. The model also could be used to identify the lacking amenities within each zone and measure the extent to which a given zone is not yet self-contained or self-sufficient. This kind of exercise would be a relevant contribution to solve mobility issues and to promote the use of non-motorized transport methods.

Our approach in this paper is static by nature; therefore, its usefulness may vary when the underlying economic and network structure change. Furthermore, since the main application of this model is intended to enhance the economic growth and the innovation through the implementation of local policies, it is necessary to better understand the behavior of the model in dynamic scenarios; namely, when many target goods are reached and a new regionalization is required to guarantee the optimality of the solution, or when the aims of the policy makers are directed to a specific set of industries rather than the ones dictated by the economic complexity. This extension is the main topic of a forthcoming paper.